\documentclass[12pt,onecolumn,amsmath,amssymb,floatix]{revtex4-1}
 
\selectfont
\usepackage{graphicx}
\usepackage{dcolumn}
\usepackage{bm}
\usepackage{hyperref}
\hypersetup{
	colorlinks=true,       
	linkcolor=red,        
	citecolor=blue,        
	urlcolor=blue,         
	pdfborder={0 0 0}      
}
\usepackage[mathlines]{lineno}
\usepackage{times}
\usepackage{float}
\usepackage{mathptmx}
\usepackage{amssymb}
\usepackage{amsmath}
\usepackage{mathtools}
\usepackage{epstopdf}
\usepackage{multirow}
\usepackage{color,booktabs}
\usepackage{longtable}
\usepackage{txfonts}
\usepackage{slashed}
\usepackage{tabularx}
\usepackage{enumitem}
\providecommand{\noopsort}
\allowdisplaybreaks

\begin{document}

\preprint{APS/123-QED}

\title{$B\to PP$ decays with Light-Cone Sum Rules}

\author{Yuanyuan Ma$^{1,}$}
\email{2022212339@institution.edu}
\author{Yan-jun Sun$^{1,}$}%
\email{sunyanjun@nwnu.edu,cn}
\affiliation{
	$^1$Institute of Theoretical Physics, College of Physics and
	Electronic Engineering, Northwest Normal University, Lanzhou 730070, China
	}

\date{\today}

\begin{abstract}

We study factorizable contributions to two-body nonleptonic decays of $B$ mesons using the light-cone sum rule method. By constructing three-point correlation functions and incorporating light-cone distribution amplitudes, we compute the hadronic matrix elements and branching ratios for decay channels $B^0 \to \pi^+\pi^-$, $B^+ \to \pi^+\pi^0$, $B^0 \to K^+\pi^-$, and $B^0 \to K^+K^-$. The results are in agreement with experimental data within uncertainties. We also estimate the branching ratio of $B_s^0 \to \pi^+\pi^-$ to be approximately $0.781\times10^{-6}$.

\end{abstract}

\maketitle


\section{\label{sec:level1}Introduction}

Heavy flavor physics is a significant branch of particle physics that focuses on the study of hadrons containing heavy quarks~\cite{Aguilar:2024ybr,ValeSilva:2024vmv,LHCb:2014iah,Belle:2014nmp,Qi:2018syl,LHCb:2014iah,Belle:2014nmp,Qi:2018syl,Brambilla:2010cs,Eichten:2017ffp,LHCb:2018vuc}. The decay of $B$ mesons involves large mixing angles in the Cabibbo-Kobayashi-Maskawa (CKM) matrix and CP violation, providing an ideal platform for testing the Standard Model (SM) and searching for potential signals of new physics (NP). Since the operation of the B factories, experimental studies of B physics have rapidly advanced. In 2003, the Belle and BaBar collaboration observed direct CP violation in the $B$ meson system~\cite{Belle:2001zzw,BaBar:2002kla}. In 2019, the LHCb collaboration reported the first observation of CP violation in the $D$ meson system~\cite{LHCb:2019hro}, offering both theoretical insight into the origin of CP violation in the SM and experimental support for possible NP effects. In recent years, the upgrades and operation of LHCb and Belle II have not only yielded high-precision experimental data, but also opened up new opportunities and challenges for theoretical research.

The decay processes of $B$ mesons are particularly important at the scale $\mu_b \approx m_b$, which is located in the transition region between the perturbative and nonperturbative regimes. Within the SM framework, the two-body nonleptonic decay amplitudes of $B$ mesons can be expressed in terms of the hadronic matrix elements $\langle M_1 M_2 | \mathcal{O}_i(\mu) | B \rangle$, where $\mathcal{O}_i$ are the effective operators in the weak Hamiltonian. However, due to the color confinement nature of Quantum Chromodynamics (QCD) and the inherently nonperturbative nature of the hadronization process, a fully self-consistent method to calculate the hadronic matrix elements $\langle M_1 M_2|\mathcal{O}_i(\mu)|B\rangle$ from first principles of QCD is still lacking. As a result, developing theoretical approaches for computing these matrix elements has become a central task. Various theoretical methods have been developed to address this challenge, including naive factorization~\cite{Wirbel:1985ji,Bauer:1984zv,Kramer:1992xr,Du:1995qy}, QCD factorization~\cite{Beneke:2000ry,Kivel:2021uzl}, perturbative QCD (pQCD)~\cite{Li:1994cka,Li:2003yj,Li:2022mtc}, soft-collinear effective theory (SCET)~\cite{Bauer:2001cu,Bauer:2001ct,Bauer:2002aj}, light-cone sum rule (LCSR)~\cite{Khodjamirian:2000mi,Wu:2002csa,Piscopo:2023opf} and so on.

In the LCSR approach, light-cone distribution amplitudes (LCDAs) were applied to handle nonperturbative effects. In Ref.~\cite{Khodjamirian:2000mi}, this method was first applied to the decay channel $\bar{B}^0_d \to \pi^+\pi^-$, and subsequent studies have extended its application to contributions from gluonic penguin operators~\cite{Khodjamirian:2002pk}, charm penguin topologies~\cite{Khodjamirian:2003eq}, and annihilation diagrams~\cite{Khodjamirian:2005wn}. Furthermore, the LCSR approach has been successfully employed in processes such as $\bar{B}_s^0 \to D_s^+\pi^-$, $\bar{B}^0 \to D^+K^-$~\cite{Piscopo:2023opf}, and $\Lambda_c^+ \to \Xi^0K^+$~\cite{Shi:2024plf}. The results show good agreement with experimental data, demonstrating the effectiveness of the LCSR method in treating non-leptonic decay processes. 

For this reason, we employ the LCSR to calculate the factorizable contributions in nonleptonic $B$ meson decays. This work will provide a baseline for subsequent inclusion of penguin diagrams and nonfactorized corrections and the exciting experiment errors require the theory to compute the tree-level contributions to a higher precision.

This paper is organized as follows. In Sec.~\ref{sec:level2}, we introduce the weak effective Hamiltonian for $B$ meson nonleptonic decays. In Sec.~\ref{sectionIII} we derive the hadronic matrix elements using the LCSR method. In Sec.~\ref{sectionIV} we present the numerical results and discussions. A brief summary is given in Sec.~\ref{sectionV}.

\section{\label{sec:level2}Weak Effective Hamiltonian}

\begin{figure}[htbp]
	\begin{center}
		\includegraphics[scale=0.48]{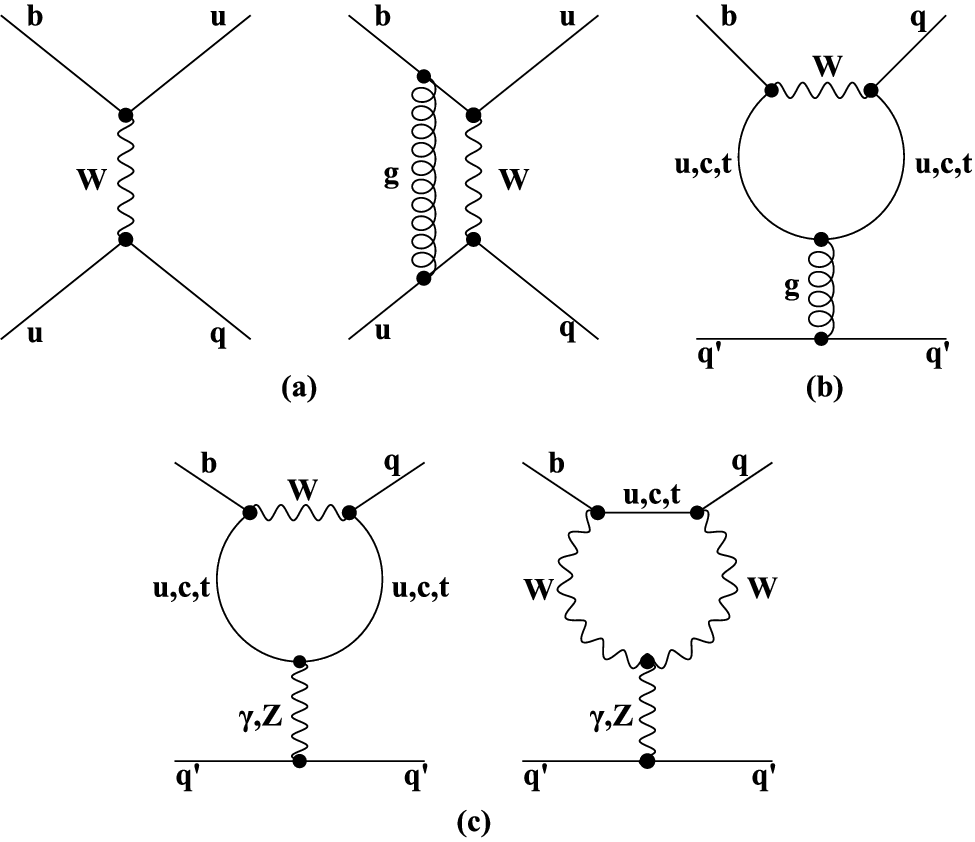}
		\caption{\label{fig1} The Feynman diagrams corresponding to the operators $\mathcal{O}_i$.}
	\end{center}
 \end{figure}

In the Standard Model, the amplitude for the exclusive nonleptonic decay $B \to P_1P_2~(B=B^0,~B^+\ \text{and}\ P_{1,2}=\pi^{\pm},~\pi^0,~K^{\pm})$ is governed by the weak effective Hamiltonian \cite{Buchalla:1995vs},

 \begin{equation}\label{Hamiltonian}
	\mathcal{H}_{\mathrm{eff}}=\frac{G_F}{\sqrt{2}}
	\lbrace V_{ub}V_{uq}^*[C_1(\mu)\mathcal{O}_1(\mu)+C_2(\mu)\mathcal{O}_2(\mu)]-V_{tb}V^*_{tq}{\sum_{i=3}^{10}}C_i(\mu)\mathcal{O}_i(\mu)+h.c.\rbrace,
 \end{equation}
where $G_F=1.166\times10^{-5}~\mathrm{GeV}^{-2}$ is the Fermi constant, $V_{ij}$ denote CKM matrix elements, and $\mathcal{O}_i~(i=1,2,3,\dots,10)$ are the corresponding four-quark operators. Specifically, $\mathcal{O}_1$ and $\mathcal{O}_2$ are the current-current operators arising from tree-level diagrams in Fig. \ref{fig1} (a), 

 \begin{equation}\label{O12}
	\mathcal{O}_1=(\bar{q}_{\alpha}u_{\beta})_{V-A}(\bar{u}_{\beta}b_{\alpha})_{V-A},\mathcal{O}_2=(\bar{q}u)_{V-A}(\bar{u}b)_{V-A}, 
 \end{equation}
where $q=d,s$, $\alpha$ and $\beta$ are color indices. $\mathcal{O}_3$-$\mathcal{O}_6$ correspond to the QCD penguin operators generated by penguin diagrams in Fig. \ref{fig1} (b),
 \begin{equation}\label{O34}
	\mathcal{O}_3=(\bar{q}b)_{V-A}{\sum_{q^\prime}}({\bar{q}^\prime}{q^\prime})_{V-A},\mathcal{O}_4=({\bar{q}_\alpha}b_\beta)_{V-A}{\sum_{q^\prime}}({\bar{q}^\prime_\beta}{q^\prime_\alpha})_{V-A},
 \end{equation}
 \begin{equation}\label{O56}
	\mathcal{O}_5=(\bar{q}b)_{V-A}{\sum_{q^\prime}}({\bar{q}^\prime}{q^\prime})_{V+A},\mathcal{O}_6=({\bar{q}_\alpha}b_\beta)_{V-A}{\sum_{q^\prime}}({\bar{q}^\prime_\beta}{q^\prime_\alpha})_{V+A}, 
 \end{equation} 
and $q^\prime=u,d,s,c,b$. Here $V \pm A$ refers to the Lorentz
structure $\gamma_{\mu}(1 \pm \gamma_5)$. In general, the contributions from the QCD penguin operators, particularly $O_3$ and $O_5$, are significantly smaller than those from the current--current operators $O_1$ and $O_2$. The contributions of the QED penguin operators $\mathcal{O}_7$ and $\mathcal{O}_8$ are relatively small and can be completely neglected. The operators $\mathcal{O}_9$ and $\mathcal{O}_{10}$ only play a role in some decay processes due to certain dynamical reasons, and their contributions can also be disregarded \cite{Buras:1998ra}. Therefore, the contributions from the QCD and QED penguin operators are not considered. We only take in account the contribution from the factorizable matrix elements of the operator $\mathcal{O}_1$ in this work.

Within the framework of Standard Model, the decay amplitude for $B \to P_1P_2$ can be expressed as $A(B \to P_1P_2)=\left\langle P_1P_2|\mathcal{H}_{\text{eff}}|B\right\rangle$. By substituting the effective Hamiltonian, the amplitude can be further expressed as 
 \begin{equation}\label{amplitude}
	A \left( B \to P_1P_2 \right) =\frac{G_F}{\sqrt{2}}V_{ub}V_{uq}^{*}( C_1+\frac{C_2}{3}) \langle P_1P_2|\mathcal{O}_1|B\rangle,
 \end{equation}
where the coefficient $(C_1+C_2/3)$ originates from the Fierz identity $O_2=1/3O_1+2\tilde{O}_1$, with the nonfactorizable contribution from the operator $\tilde{O}_1$ being negligible~\cite{Wu:2002csa}.

And the decay branching ratio and decay width are  expressed as:

 \begin{equation}\label{BR}
	\mathcal{B} \left( B\rightarrow P_1P_2 \right) =\tau_{B}\frac{|p_c|}{8\pi m_{B}^{2}}|A (B\rightarrow P_1P_2 )|^2,
 \end{equation}
 \begin{equation}\label{width}
 	\Gamma \left( B\rightarrow P_1P_2 \right) =\frac{|p_c|}{8\pi m_{B}^{2}}|A (B\rightarrow P_1P_2 )|^2,
 \end{equation} 
where $\tau_{B}$ is lifetime of $B$ meson, $|p_c|$ represents the momentum of the final-state pseudoscalar meson in the rest frame of the $B$ meson :
 \begin{equation}\label{momentum}
	|p_c|=\frac{1}{2m_B}\sqrt{[m_B^2-(m_{P_1}+ m_{P_2})^2][m_B^2-(m_{P_1}-m_{P_2})^2]}.
 \end{equation}


\section{\label{sectionIII}Hadronic matrix elements with LCSRs}


The hadronic matrix element $\langle P_1P_2|\mathcal{O}_1|B\rangle$ is a key object for the decay amplitudes of  process $B \to P_1P_2$. To obtain this matrix element, we construct a three-point correlation function~\cite{Khodjamirian:2000mi}

 \begin{equation}\label{corr}
 	F^{(\mathcal{O}_1)}_{\alpha}(p,q,k)=\mathrm{i}\int{\mathrm{d}^4x}\ \mathrm{e}^{-\mathrm{i}(p-q) \cdot x}\
 	\mathrm{i}\int{\mathrm{d}^4y}\ \mathrm{e}^{\mathrm{i}(p-k) \cdot y}
 	\langle0|T\lbrace j_{\alpha5}^{(P_2)}(y)\mathcal{O}_1(0)j_{5}^{(B)}(x)\rbrace|P_1(q)\rangle, 
 \end{equation}
where $j_{\alpha5}^{(P_2)}=\bar{u}\gamma_{\alpha}\gamma_5q~(q=u,d,s)$ and $j^{(B)}_5=\bar{b}\mathrm{i}\gamma_5q^\prime~(q^\prime=u,d)$ denote the interpolating currents for the pseudoscalar meson $P_2$ and the $B$ meson, respectively. In dealing with the correlation function, in order to unambiguously extract a matrix element $\langle 0|j_{\alpha5}^{(P_2)}| P_2\rangle $, we temporarily shift a pseudoscalar meson $P_1$ from the final state to the initial state in the matrix element, obeying on-shell condition $q^2=m_{P_1}^2$. We take the chiral limit $m_{P_2}=0$, except in cases where chirally enhanced terms such as $m_{P_2}^2/{(m_u+m_q)}$ appear. 

The decomposition of the correlation function Eq. \eqref{corr} in terms of independent momenta is straightforward and involves four invariant amplitudes:
 \begin{equation}\label{Invariant Amplitude}
	F^{(\mathcal{O}_1)}_{\alpha}(p,q,k)=(p-k)_\alpha F^{(\mathcal{O}_1)}+q_\alpha \tilde{F}_1^{(\mathcal{O}_1)}+k_\alpha \tilde{F}_2^{(\mathcal{O}_1)}+\epsilon_{\alpha\beta\lambda\rho}q^{\beta}p^{\lambda}k^{\rho}\tilde{F}_3^{(\mathcal{O}_1)}, 
 \end{equation}
with only $F^{(\mathcal{O}_1)}$ being relevant to the present work. Due to the absence of external momentum flowing out of the weak operator vertex, the four-momentum of the $b$ quark remains unchanged before and after the decay. To address this, an auxiliary momentum $k$ is artificially introduced, such that the momentum of the $B$ meson after decay becomes $P = p - k - q$. When $k \neq 0$, the invariant amplitude $F^{(\mathcal{O}_1)}((p-k)^2,(p-q)^2,P^2)$ includes contributions from excited states and the continuum spectrum of the $B$ meson. Since the auxiliary momentum $k$ does not correspond to any physical external state, it vanishes in the evaluation of the ground-state contribution to the decay process $B \to P_1 P_2$~\cite{Khodjamirian:2000mi}.

At the quark-gluon level, we expand the correlation function Eq. \eqref{corr} using Wick's theorem. We take the decay channel $B^0 \to \pi^+ \pi^-$ as an example and choose $q = d$ , $q^\prime = d$. The time-ordered product of the two current operators and the four-quark operator in the correlator is expanded near the light-cone region, where $x^2 \sim y^2 \sim (x-y)^2 \sim 0$. To ensure the validity of this expansion, one must choose an appropriate kinematic region such that the external momentum squares $(p-q)^2$,~$(p-k)^2$, and the kinematic invariant $P^2$ are all spacelike and sufficiently large. To simplify calculations, we set external momentum squared and the kinematical invariant $k^2=p^2=0$, respectively. In summary, the chosen kinematic region is:

\begin{equation}\label{kinematic region}
	q^2=p^2=k^2=0,~|(p - k)^2| \sim |(p - q)^2| \sim |P^2| \gg \Lambda_{\text{QCD}}^2.
\end{equation}
The correlation function Eq.~\eqref{corr} becomes:
 \begin{equation}\label{Wick}
	\begin{aligned}
		F_{\alpha}^{(\mathcal{O}_1)}(p,q,k)
		=&\mathrm{i}^3\int{\mathrm{d}^4x}\ \mathrm{e}^{-\mathrm{i}(p-q) \cdot x}\int{\mathrm{d}^4y}\ \mathrm{e}^{\mathrm{i}(p-k) \cdot y}
		\langle0|T\lbrace\bar{u}(y)\gamma_{\alpha}\gamma_5q(y)\bar{q}(0)\gamma_{\mu}\gamma_5u(0)\bar{u}(0)\gamma^{\mu}b(0)\bar{b}(x)\gamma_5q^\prime(x)\rbrace|P_1(q)\rangle\\
		=&-\mathrm{i}^3\int{\mathrm{d}^4x}\ \mathrm{e}^{-\mathrm{i}(p-q) \cdot x}\int{\mathrm{d}^4y}\ \mathrm{e}^{\mathrm{i}(p-k) \cdot y}\langle 0|:\bar{u}(0)\gamma^{\mu}\overbracket{b(0)\bar{b}}(x)\overbracket{d(y)\bar{d}}(0)\gamma_{\mu}\gamma_5\overbracket{u(0)\bar{u}}(y)\gamma_{\alpha}\gamma_5\gamma_5d(x):| P_1(q)\rangle\\
		&+\cdots,
	\end{aligned}
 \end{equation}
where all possible contractions of quark fields are expressed in terms of free quark propagators, such as $b(0)\bar{b}(x)$, $u(0)\bar{u}(y)$, and so on. The free quark propagator is given by
 \begin{equation}\label{propagator}
	\mathrm{i}S_F^0(x,0)=\mathrm{i}\int{\frac{\mathrm{d}^4p}{(2\pi)^4}\ \mathrm{e}^{-\mathrm{i}p\cdot x}}\ \frac{m+\slashed{p}}{m^2-p^2}.
 \end{equation}
The remaining uncontracted part, such as $\langle0|\bar{u}(0)\Gamma_i q(x)|P_1\rangle$, are expressed in terms of the LCDAs of the pseudoscalar meson~\cite{Belyaev:1994zk},
\begin{equation}\label{LCDAs2}
	\langle 0|\bar{u}\left( 0 \right) \gamma _{\alpha}\gamma _5d\left( x \right) |P\left( q \right) \rangle
	=\mathrm{i}q_{\alpha}f_{P}\int_0^1{\mathrm{d}u}\ \mathrm{e}^{-\mathrm{i}uq\cdot x}\varphi _{2}\left( u,\mu \right),
\end{equation}
\begin{equation}\label{LCDAs3}
	\langle 0|\bar{u}\left( 0 \right) \mathrm{i}\gamma _5q\left( x \right) |P\left( q \right) \rangle
	=\frac{m_{P}^{2}f_{P}}{m_u+m_q}\int_0^1{\mathrm{d}u}\ \mathrm{e}^{-\mathrm{i}uq\cdot x}\varphi _{3}^P\left( u,\mu \right),
\end{equation}
where $\varphi_2(u,\mu)$ and $\varphi_3^P(u,\mu)$ denote the twist-2 and twist-3 distribution amplitudes of the pseudoscalar meson, respectively, and their Gegenbauer expansions are listed in the Appendix.

The invariant amplitude $F^{(\mathcal{O}_1)}_{\mathrm{QCD}}$ can be expressed using dispersion relations:
 \begin{equation}\label{dispersion relations}
	F^{(\mathcal{O}_1)}_{\text{QCD}}((p-k)^2,(p-q)^2,P^2)=\frac{1}{\pi}\int_{0}^{\infty}\mathrm{d}s\ \frac{\mathrm{Im}F^{(\mathcal{O}_1)}_{\text{QCD}}(s,(p-q)^2,P^2)}{s-(p-k)^2}. 
 \end{equation}
By inserting a complete set of intermediate states with the same quantum numbers of the final-state pseudoscalar meson into Eq. \eqref{corr}, we obtain hadronic representation of the correlation function:
 \begin{equation}\label{Hadr}
	F^{(\mathcal{O}_1)}((p-k)^2,(p-q)^2,P^2)=
	\frac{\mathrm{i}f_{P_2}\Pi_{PP}^{(\mathcal{O}_1)}((p-q)^2,P^2)}{-(p-k)^2}
	+\int_{s_0}^{\infty}\mathrm{d}s\ \frac{\rho_h(s,(p-q)^2,P^2)}{s-(p-k)^2}, 
 \end{equation}
where decay constant of pseudoscalar meson is defined as $\left\langle 0\left| \bar{u}\gamma_{\alpha}\gamma_5q\right| P_2(p-k)\right\rangle=i(p-k)_\alpha f_{P_2}$, $s_0$
is the threshold, the spectral function $\rho_h$ represents the contributions of excited states and continuum in the pseudoscalar meson channel. Furthermore, the integral over $\rho_h$ can be approximated using quark-hadron duality,
\begin{equation}\label{QH duality}
	\int_{s_0}^{\infty}{\mathrm{d}s}\ \frac{\rho_h(s,(p-q^2),P^2)}{s-(p-k)^2}=
	\frac{1}{\pi}\int_{s_0}^{\infty}{\mathrm{d}s}\ \frac{\mathrm{Im}F_\text{QCD}^{(\mathcal{O}_1)}(s,(p-q)^2,P^2)}{s-(p-k)^2}.
\end{equation}
And the hadronic matrix element in Eq.~\eqref{Hadr} is
 \begin{equation}\label{two point corr}
	\Pi_{PP}^{(\mathcal{O}_1)}((p-q)^2,P^2)=\mathrm{i}\int{\mathrm{d}^4x}\ \mathrm{e}^{-\mathrm{i}(p-q) \cdot x}\langle P_2(p-k)| T\lbrace {\mathcal{O}_1(0)j_5^{(B)}(x)}\rbrace |P_1(q)\rangle.  
 \end{equation}
By combining Eqs.~\eqref{dispersion relations}~$-$~\eqref{QH duality}, and performing Borel transform with respect to the variable $(p-k)^2$, we transform the Eq.~\eqref{two point corr} into
 \begin{equation}\label{DR}
	\Pi_{PP}^{(\mathcal{O}_1)}((p-q)^2,P^2)=
	-\frac{\mathrm{i}}{\pi f_{P_2}}\int_{0}^{s_0}\mathrm{d}s\  \mathrm{e}^{-\frac{s}{M_1^2}}\mathrm{Im}F_\text{QCD}^{(\mathcal{O}_1)}(s,(p-q)^2,P^2). 
 \end{equation} 

To obtain the hadronic matrix element for the nonleptonic decay of the $B$ meson, we continue the two-point correlation function Eq.~\eqref{two point corr} to large spacelike region,
 \begin{equation}\label{AC corr}
	\Pi_{PP}^{(\mathcal{O}_1)}((p-q)^2,P^2)=\mathrm{i}\int{\mathrm{d}^4x}\ \mathrm{e}^{-\mathrm{i}(p-q) \cdot x}\langle P_2(p-k)P_1(-q)| T\lbrace {\mathcal{O}_1(0)j_5^{(B)}(x)}\rbrace|0\rangle,
 \end{equation}
with $P^2 = m_B^2$. Inserting a complete set of states with the same quantum numbers as the $B$ meson between the operators $\mathcal{O}_1$ and $j_5^{(B)}$, we obtain hadronic representation of Eq.~\eqref{AC corr}
 \begin{equation}\label{ME}
	\Pi_{PP}^{(\mathcal{O}_1)}((p-q)^2,m_B^2)=
	\frac{(m_B^2/m_b)f_B}{m_B^2-(p-q)^2}
	\left\langle P_2(p-k)P_1(-q)\left| \mathcal{O}_1\right| B(p-q)\right\rangle 
	+\int_{s_0^B}^{\infty}\mathrm{d}s^\prime\frac{\rho^{\prime}(s^\prime,m_B^2)}{s^\prime-(p-q)^2}, 
 \end{equation} 
where $\langle B|j_5^{(B)}|0\rangle =m_B^2f_B/m_b$ denotes the decay constant of the $B$ meson, and the spectral function $\rho^\prime$ is the contributions of excited states and continuum in the $B$ meson channel. Similarly to the derivation of Eq.~\eqref{DR}, based on quark-hadron duality, the two-point correlation Eq.~\eqref{AC corr} function can be processed accordingly. Note that in the ground-state contribution the auxiliary momentum $k$ vanishes, due to simultaneous conditions $(p - q - k)^2 = m_B^2$ and $(p - q)^2 = m_B^2$, so that matrix element $\left\langle P_2(p)P_1(-q)\left| \mathcal{O}_1\right| B(p-q)\right\rangle$ is recovered~\cite{Khodjamirian:2000mi}. Substituting the Eq.~\eqref{DR} into the left hand of the Eq.~\eqref{ME} and Borel transforming $(p-q)^2 \to M_2^2$, we obtain the decay amplitude for $B \to P_1P_2$,
 \begin{equation}\label{matrix element}
	\left\langle P_2(p)P_1(-q)\left| \mathcal{O}_1\right| B(p-q)\right\rangle=
	\frac{-\mathrm{i}m_b}{\pi^2 m_B^2 f_{P_2} f_B}\int_{0}^{s_0}\mathrm{d}s\ \mathrm{e}^{-\frac{s}{M_1^2}}\int_{m_b^2}^{s_0^B}\mathrm{d}s^\prime\ \mathrm{e}^{\frac{m_B^2-s^\prime}{M_2^2}} \mathrm{Im}_{s}\mathrm{Im}_{s^{\prime}} F_{\mathrm{QCD}}^{(\mathcal{O}_1)}(s,s^\prime,m_B^2), 
 \end{equation} 
where $s_0^B$ is threshold of $B$ meson channel. Here, taking $B^0 \to \pi^+ \pi^-$ as an example, we present the explicit form of $\mathrm{Im}_{s}\mathrm{Im}_{s^{\prime}} F_{\mathrm{QCD}}^{(\mathcal{O}_1)}(s,s^\prime,m_B^2)$,

 \begin{equation}\label{3.25}
	\begin{aligned}
		{\mathrm{Im}_{s}\mathrm{Im}_{s^\prime}F_\mathrm{QCD}^{(\mathcal{O}_1)}}=&-\frac{m_bf_{\pi^-}(m_{B}^2-2m_{\pi^-}^2)}
		{96\pi}\int_0^1{\mathrm{d}u}
		\frac{1}{(1-u)M_2^2}\ \mathrm{e}^{-\frac{m_b^2-u^2m_{\pi^-}^2}{(1-u)M_2^2}}
		\varphi_{2}(u)\\
		&-\frac{f_{\pi^-}m_{B}^2m_{\pi^-}^2}{48\pi(m_u+m_d)}\int_0^1{\mathrm{d}u}
		\frac{1}{(1-u)M_2^2}\ \mathrm{e}^{-\frac{m_b^2-u^2m_{\pi^-}^2}{(1-u)M_2^2}}
		\varphi_{3}^P(u).
	\end{aligned}
 \end{equation}

Matching the results from the OPE side and the hadronic spectral representation, we obtain the hadronic matrix element for $B \to \pi\pi$, $B \to K\pi$, and $B \to KK$ as follows:

 \begin{equation}
	\begin{aligned}
		\langle\pi^-\pi^+|\mathcal{O}_1|B^0\rangle = &\frac{\mathrm{i}m_b}{48\pi^3f_{B^0}} \int_0^{s_0^\pi} \mathrm{d}s\ \mathrm{e}^{-\frac{s}{M_1^2}} \int_{m_b^2}^{s_0^B} \mathrm{d}s^\prime \mathrm{e}^{\frac{m_B^2-s^\prime}{M_2^2}} \int_0^1 \mathrm{d}u\ \frac{1}{(1-u)M_2^2}\mathrm{e}^{-\frac{m_b^2-u^2m_\pi^2}{(1-u)M_2^2}} \\
		&\times \bigg[ \frac{m_b^2(m_B^2-2m_\pi^2)}{2m_B^2} \varphi_{2;\pi}(u) + \frac{m_\pi^2m_b}{m_u+m_d} \left(1 + \frac{u(m_B^2-2m_\pi^2)}{m_B^2}\right) \varphi_{3;\pi}^{P}(u) \bigg],
	\end{aligned}
 \end{equation}

 \begin{equation}
	\begin{aligned}
		\langle \pi^0 \pi^+ | \mathcal{O}_1 | B^+ \rangle 
		&= \frac{\mathrm{i}\sqrt{2}m_b}{192\pi^3 f_{B^+} m_B^2} \int_0^{s_0^\pi} \mathrm{d}s\ \mathrm{e}^{-\frac{s}{M_1^2}} \int_{m_b^2}^{s_0^B} \mathrm{d}s' \mathrm{e}^{\frac{m_B^2 - s'}{M_2^2}} \int_0^1 \mathrm{d}u\ \frac{1}{(1-u)M_2^2}\mathrm{e}^{-\frac{m_b^2 - u^2 m_{\pi^+}^2 - u m_{\pi^0}^2 + u m_{\pi^+}^2}{(1-u)M_2^2}} \\
		&\quad \times \bigg[ m_b \left( m_B^2 - m_{\pi^+}^2 - m_{\pi^0}^2 \right) \varphi_{2;\pi}(u) - \frac{m_{\pi^+}^2}{m_u + m_d} \left( \left( m_B^2 + m_{\pi^0}^2 - m_{\pi^+}^2 \right) - u \left( m_B^2 - m_{\pi^+}^2 - m_{\pi^0}^2 \right) \right) \varphi_{3;\pi}^{P}(u) \bigg],
	\end{aligned}
 \end{equation}
 
 \begin{equation}
 	\begin{aligned}
 		\langle \pi^- K^+ | \mathcal{O}_1 | B^0 \rangle 
 		=& \frac{\mathrm{i} (m_B^2 - m_{\pi^-}^2 - m_{K^+}^2)}{48\pi^3 f_{B^0} m_B^2} 
 		\int_0^{s_0^K} \mathrm{d}s\ \mathrm{e}^{-\frac{s}{M_1^2}} 
 		\int_{m_b^2}^{s_0^B} \mathrm{d}s^{\prime}\ \mathrm{e}^{\frac{m_B^2 - s^{\prime}}{M_2^2}} \int_0^1 \mathrm{d}u\ \frac{1}{(1-u) M_2^2}\mathrm{e}^{-\frac{m_b^2 - u^2 m_{\pi^-}^2 - u m_{K^+}^2 + u m_{\pi^-}^2}{(1-u) M_2^2}} \\
 		&\times \left[ \frac{m_b^2}{2} \varphi_{2;\pi}(u) + \frac{m_b (u m_{\pi^-}^2 - m_{K^+}^2)}{m_u + m_s} \varphi_{3;K}^{P}(u) \right],
 	\end{aligned}
 \end{equation}
 
 \begin{equation}
 	\begin{aligned}
 		\langle K^- K^+ | \mathcal{O}_1 | B^0 \rangle 
 		&= \frac{\mathrm{i}}{48\pi^3 f_{B^0} m_B^2} 
 		\int_0^{s_0^K} \mathrm{d}s\ \mathrm{e}^{-\frac{s}{M_1^2}} 
 		\int_{m_b^2}^{s_0^B} \mathrm{d}s^{\prime}\ \mathrm{e}^{\frac{m_B^2 - s^{\prime}}{M_2^2}} 
 		\int_0^1 \mathrm{d}u\ \frac{1}{(1-u) M_2^2}\mathrm{e}^{-\frac{m_b^2 - u^2 m_K^2}{(1-u) M_2^2}} \\
 		&\quad \times \left[ \frac{m_b^2 (m_B^2 - 2 m_K^2)}{2m_B^2} \varphi_{2;K}(u) 
 		+ \frac{m_K^2 m_b}{m_u + m_s}(1 + u \frac{m_B^2 - 2 m_K^2}{m_B^2}) \varphi_{3;K}^{P}(u) \right].
 	\end{aligned}
 \end{equation}

\section{\label{sectionIV}Numerical results}

\subsection{Input Parameters}

The input parameters employed in our analysis, including quark masses, meson masses, lifetimes, and related quantities (all evaluated at the energy scale of 2.4~GeV), are summarized in Table~\ref{tab:mass}.

 \renewcommand{\tabcolsep}{0.1cm}
 \renewcommand{\arraystretch}{1.5}
 \begin{table}[H]
	\caption{\label{tab:mass}
		The input parameters for theoretical calculations, including quark masses, meson masses, lifetimes \cite{ParticleDataGroup:2024cfk}, decay constants \cite{Arifi:2022pal}, CKM matrix elements \cite{ParticleDataGroup:2024cfk}, Wilson coefficients \cite{Buchalla:1995vs}, and Gegenbauer coefficients of the LCDAs for the pion and kaon ($\mu_b=2.4\ \mathrm{GeV}$) \cite{Shi:2024plf,Belyaev:1994zk,Ball:2004ye}.}
	\renewcommand{\arraystretch}{1.5}
		\begin{tabular}{lllll}
			\bottomrule[1.0pt]\bottomrule[0.5pt]
			Quark masses &$m_b=4.183\ \mathrm{GeV}$ &$m_s=0.093\ \mathrm{GeV}$ &$m_d=4.70\ \mathrm{MeV}$ &$m_u=2.16\ \mathrm{MeV}$\\
			Meson masses &$m_{\pi^\pm}=139.57\ \mathrm{MeV}$ &$m_{\pi^0}=134.98\ \mathrm{MeV}$ &$m_{K^\pm}=493.68\ \mathrm{MeV}$ &$m_{B^{+/0}}=5.28\ \mathrm{GeV}$\\
			Meson lifetimes &$\tau_{B^+}=1.638\ \mathrm{fs}$ &$\tau_{B^0}=1.517\ \mathrm{fs}$ & &\\
			Meson decay constants &$f_{B^+}=188\ \mathrm{MeV}$ &$f_{B^0}=190\ \mathrm{MeV}$ & &\\
			Wilson coefficients  &$C_1=1.185$ &$C_2=-0.387$ & &\\
			CKM matrix elements &$V_{ub}=0.00382$ &$V_{ud}=0.97367$ &$V_{us}=0.22431$ &\\
			Gegenbauer coefficients  &$a_{2}^{\pi}(\mu_b)=0.35$ &$a_{4}^{\pi}(\mu_b)=0.18$ &$B_{2}^{\pi}(\mu_b)=0.29$ &$B_{4}^{\pi}(\mu_b)=0.58$\\
			&$a_{1}^{K}(\mu _b)=0.156$ &$a_{2}^{K}(\mu _b)=0.181$ &$a_{3}^{K}(\mu _b)=0.056$ &$a_{4}^{K}(\mu _b)=0.061$ \\
			&$\eta_3^K(\mu _b)=0.009$ &$\omega_3^K(\mu _b)=-2.771$ &$\rho_3^K(\mu _b)=0.309$  &\\
			\bottomrule[0.5pt]\bottomrule[1.0pt] 
		\end{tabular}
 \end{table}


In our LCSR methods, the results depend on the chosen threshold parameters \( s_0^\pi \) and \( s_0^K \), as well as on the Borel parameters \( M_1^2 \) and \( M_2^2 \), which are introduced to suppress contributions from higher resonances and the continuum. The threshold parameters $s_{0}^{\pi}$ and $s_{0}^{K}$ are set equal to the squared masses of the first excited states of pion and kaon, respectively. For the $B$ meson, the threshold parameter is taken as $s_0^B=35\ \mathrm{GeV}^2$ from Ref.~\cite{Khodjamirian:2000mi}. These parameters, together with the Borel parameters $M_1^2$ and $M_2^2$, are presented in Table~\ref{tab:Borel}.

 \renewcommand{\tabcolsep}{0.35cm}
 \renewcommand{\arraystretch}{1.5}
 \begin{table}[H]
	\caption{The threshold and Borel parameters for the $B \to P_1P_2$ decay channels.}
	\centering
	\begin{tabular}{ccc}
		\bottomrule[1.0pt]\bottomrule[0.5pt]
		Decay channels        &Threshold parameters  &Borel parameters \\
		\hline
		$B^0\to\pi^+\pi^-$    &\multirow{2}{*}{$s_{0}^{\pi}=1.69\ \mathrm{GeV}^2$,$s_{0}^{B}=35\ \mathrm{GeV}^2$}    &$M_{1}^{2}=0.5\pm0.1\ \mathrm{GeV}^2, M_{2}^{2}=8\pm1\ \mathrm{GeV}^2$       \\
		$B^+\to\pi^+\pi^0$    &   &$M_{1}^{2}=0.5\pm0.1\ \mathrm{GeV}^2, M_{2}^{2}=7\pm1\ \mathrm{GeV}^2$      \\
		$B^0\to K^+\pi^-$     &\multirow{2}{*}{$s_{0}^{K}=2.13\ \mathrm{GeV}^2$,$s_{0}^{B}=35\ \mathrm{GeV}^2$}   &$M_{1}^{2}=2\pm1\ \mathrm{GeV}^2, M_{2}^{2}=7\pm1\ \mathrm{GeV}^2$       \\
		$B^0\to K^+K^-$       &   &$M_{1}^{2}=5\pm1\ \mathrm{GeV}^2, M_{2}^{2}=8\pm1\ \mathrm{GeV}^2$       \\
		\bottomrule[0.5pt]\bottomrule[1.0pt]
	\end{tabular}\label{tab:Borel}
 \end{table}

\begin{figure}[H]
	\begin{center}
		\includegraphics[scale=0.3]{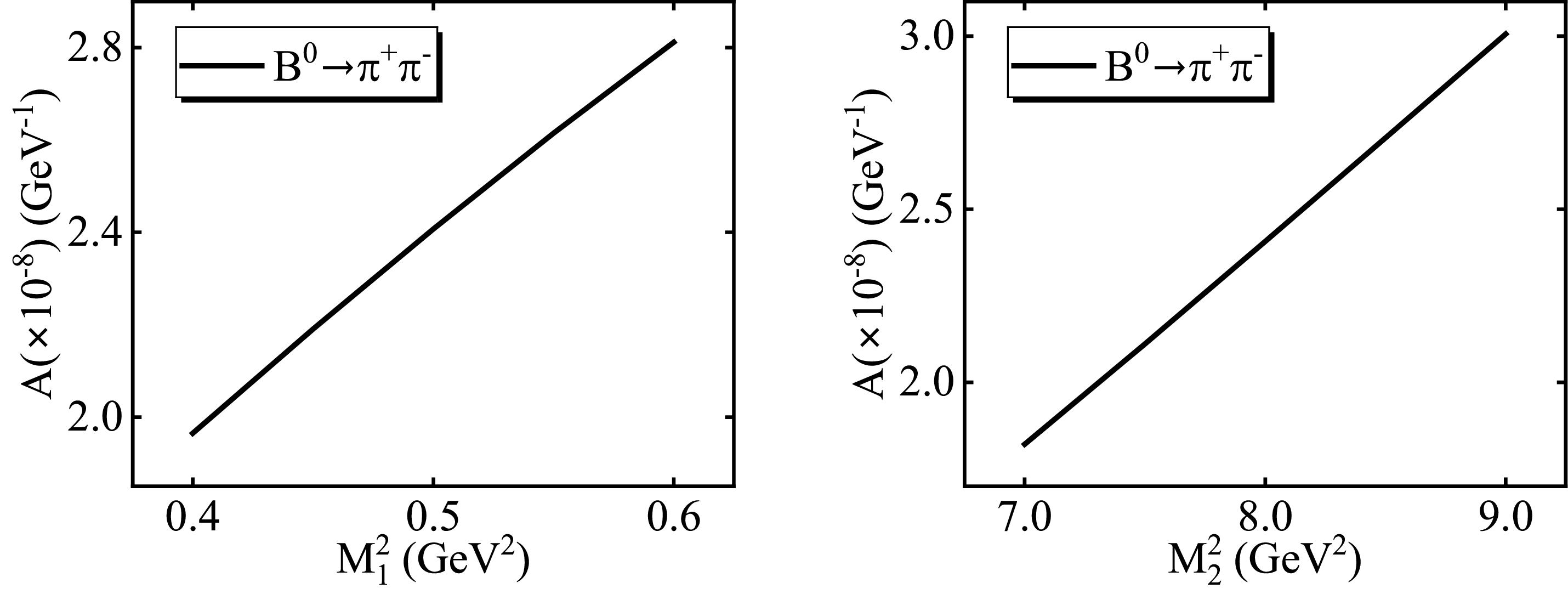}
		\caption{\label{B0pipi} Dependence of the decay amplitudes on the Borel parameters $M_1^2$ and $M_2^2$ for $B^0\to\pi^+\pi^-$.}
	\end{center}
 \end{figure}

\begin{figure}[H]
	\begin{center}
		\includegraphics[scale=0.3]{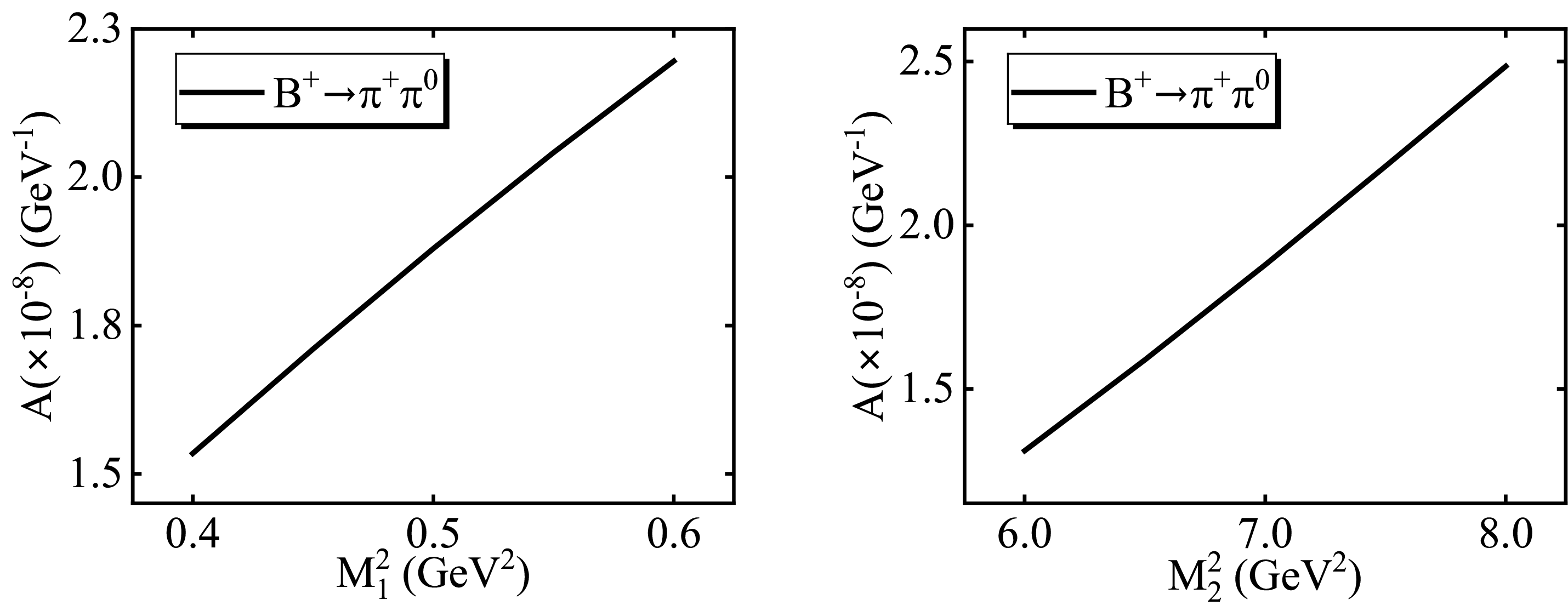}
		\caption{\label{B+pipi} Dependence of the decay amplitudes on the Borel parameters $M_1^2$ and $M_2^2$ for $B^+\to\pi^+\pi^0$.}
	\end{center}
 \end{figure}

\begin{figure}[H]
	\begin{center}
		\includegraphics[scale=0.3]{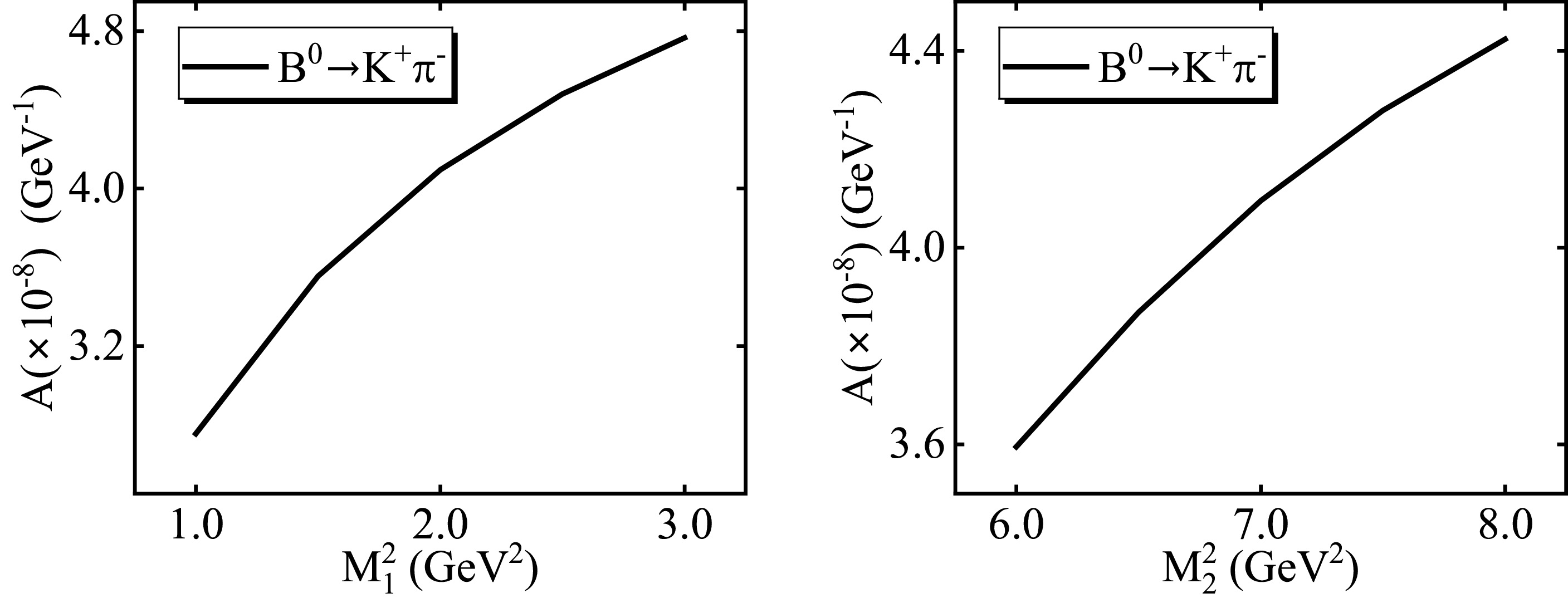}
		\caption{\label{B0Kpi} Dependence of the decay amplitudes on the Borel parameters $M_1^2$ and $M_2^2$ for $B^0\to K^+\pi^-$.}
	\end{center}
 \end{figure}

\begin{figure}[H]
	\begin{center}
		\includegraphics[scale=0.3]{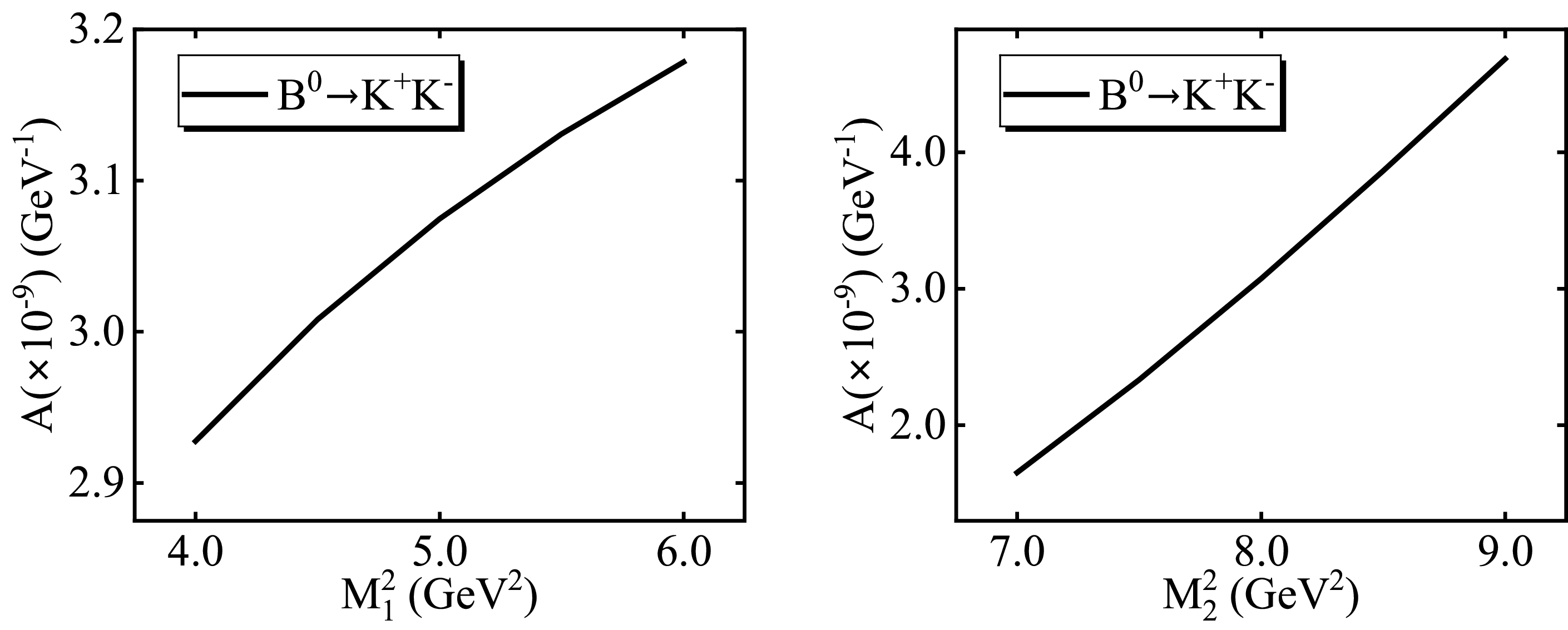}
		\caption{\label{B0KK} Dependence of the decay amplitudes on the Borel parameters $M_1^2$ and $M_2^2$ for $B^0\to K^+K^-$.}
	\end{center}
 \end{figure}

We then discuss the dependence of the numerical results on the Borel parameters. Figs.~\ref{B0pipi}$-$\ref{B0KK} illustrate the dependence of the decay amplitudes on Borel parameters \( M_1^2 \) and \( M_2^2 \) for the four decay channels: \( B^0 \to \pi^+ \pi^- \), \( B^+ \to \pi^+ \pi^0 \), \( B^0 \to K^+ \pi^- \), and \( B^0 \to K^+ K^- \), respectively.

As shown in the figures, the decay amplitudes~\(A\)~for the processes $B \to P_1P_2$ are of the order of \(10^{-8}\) or \(10^{-9}\), and exhibit mild dependence on the Borel parameters. 

\subsection{Numerical Results and Discussions}

Based on the relation between the hadronic matrix element and the decay amplitude given in Eq.~(\ref{amplitude}), we obtain the decay amplitudes and widths for the processes $B \to P_1P_2$, with the results summarized in Table~\ref{tab:Amplitude and width}.



 \renewcommand{\tabcolsep}{0.30cm}
 \renewcommand{\arraystretch}{1.5}
 \begin{table}[H]
	\caption{The decay amplitudes and widths for $B \to P_1P_2$.}
	\centering
	\begin{tabular}{ccc}
		\bottomrule[1.0pt]\bottomrule[0.5pt]
		Decay channels        &Decay Amplitudes($\mathrm{GeV}^{-1}$)  &Decay Widths($\mathrm{GeV}$) \\
		\hline
		$B^0\to\pi^+\pi^-$    &$2.41\times10^{-8}$    &$2.18\times10^{-18}$  \\
		$B^+\to\pi^+\pi^0$    &$1.87\times10^{-8}$    &$1.88\times10^{-18}$  \\
		$B^0\to K^+\pi^-$     &$4.10\times10^{-8}$    &$8.85\times10^{-18}$  \\
		$B^0\to K^+K^-$       &$3.07\times10^{-9}$    &$3.42\times10^{-20}$  \\
		\bottomrule[0.5pt]\bottomrule[1.0pt]
   \end{tabular}\label{tab:Amplitude and width}
 \end{table}

Furthermore, we obtain the branching ratios for these decay channels and compare them with those obtained from other approaches, such as pQCD~\cite{Chai:2022ptk}, QCDF~\cite{Cheng:2009cn}, and LCSR~\cite{Khodjamirian:2005wn}, as summarized in Table~\ref{tab:BR}. The first error reflects the variation of the Borel parameter $M_1^2$, and the second is due to that of $M_2^2$. Our results exhibit reasonable agreement with the experimental measurements~\cite{Belle-II:2023ksq,LHCb:2016inp}. While minor discrepancies are found in some decay modes when compared with predictions from other theoretical approaches, the overall consistency supports the reliability of estimating factorizable contributions to hadronic $B$-meson decays within the LCSR framework.

 
 \renewcommand{\tabcolsep}{0.09cm}
 \renewcommand{\arraystretch}{1.5}
 \begin{table}[H]
	\caption{The branching ratios for $B \to P_1P_2$ in our LCSR approach (in units of $10^{-6}$) and comparisons with other theoretical and experimental results.}
	\centering
	\begin{tabular}{cccccc}
		\bottomrule[1.0pt]\bottomrule[0.5pt]
		Decay channels        &This work  &pQCD\cite{Chai:2022ptk} &QCDF\cite{Cheng:2009cn} &LCSR\cite{Khodjamirian:2005wn} &Expt.\cite{Belle-II:2023ksq,LHCb:2016inp}\\
		\hline
		$B^0\to\pi^+\pi^-$    &$5.01_{-1.67-2.14}^{+1.82+2.80}$   &$7.31_{-1.68-0.36}^{+2.35+0.38}$      &$7.0_{-1.0}^{+0.8}$     &$6.7^{+1.8+0.9}_{-1.5-0.8}$ &$5.83\pm0.22\pm0.17$  \\
		$B^+\to\pi^+\pi^0$    &$4.67_{-1.55-2.39}^{+1.70+3.49}$   &$4.18_{-0.94-0.22}^{+1.30+0.22}$      &$5.9_{-1.6}^{+2.6}$     &$9.7^{+2.3+1.2}_{-1.9-1.2}$ &$5.10\pm0.29\pm0.27$  \\
		$B^0\to K^+\pi^-$     &$20.39_{-11.17-4.67}^{+7.22+3.39}$  &$17.1_{-3.7-0.1}^{+5.2+0.1}$          &$19.3_{-7.8}^{+11.4}$   &$-$ &$20.67\pm0.37\pm0.62$  \\
		$B^0\to K^+K^-$       &$0.079_{-0.005-0.056}^{+0.007+0.078}$  &$0.046_{-0.039-0.008}^{+0.058+0.009}$ &$0.1\pm0.04$            &$-$ &$0.078\pm0.012\pm0.008\pm0.002$  \\
		\bottomrule[0.5pt]\bottomrule[1.0pt]
	\end{tabular}\label{tab:BR}
 \end{table}

Table~\ref{tab:BR} shows that for the first two decay channels, $B^0 \to \pi^+ \pi^-$ and $B^+ \to \pi^+ \pi^0$, our calculated branching ratios are slightly smaller than the experimental results. This is because our calculations include only the tree-level contributions to these decay channels, and neglect the corrections from both soft and hard gluon exchanges. 


Based on the experimentally measured branching ratios for $B^0 \to \pi^+ \pi^-$ and $B^0 \to K^+ \pi^-$, we extract the numerical values of the CKM matrix element combinations $|V_{ub}V_{ud}^*|$ and $|V_{ub}V_{us}^*|$, and evaluate their ratio to be approximately 0.18. Since our calculation does not include the penguin contributions to the $B^0 \to K^+ \pi^-$ decay, the extracted ratio is slightly smaller than the PDG average value $|V_{us}/V_{ud}| \approx 0.23$~\cite{ParticleDataGroup:2024cfk}. According to SU(3) flavor symmetry, the amplitude for $B^0 \to K^+ \pi^-$ decay receives significant contributions from both tree-level and penguin topologies~\cite{Gronau:1994rj}, in contrast to the $B^0 \to \pi^+ \pi^-$ decay, which is predominantly governed by tree-level dynamics.


Using the above branching ratios, we can further estimate the branching ratio for $B_s^0 \to \pi^+ \pi^-$ decay. Ref.~\cite{LHCb:2016inp} reported the following numerical result:
 \begin{equation}\label{Bs}
	\frac{f_s}{f_d} \frac{\mathcal{B}(B_s^0 \to \pi^+\pi^-)}{\mathcal{B}(B^0 \to K^+\pi^-)} = (9.15 \pm 0.71 \pm 0.83) \times 10^{-3},
 \end{equation}
where $f_s$ and $f_d$ denote the probabilities for the hadronization of $b$-quark into $B_s^0$ and $B^0$ meson, respectively. By substituting $\mathcal{B}(B^0 \to K^+\pi^-)$ obtained from our calculation into the above Eq.~\eqref{Bs}, and using the value $f_s/f_d = 0.256 \pm 0.020$ from Ref.~\cite{LHCb:2013vfg}, we estimate:
 \begin{equation}
	\mathcal{B}(B_s^0 \to \pi^+\pi^-) \approx (0.781 \pm 0.233)\times10^{-6},
 \end{equation}
with the error originates from the assumption of SU(3)  flavor  symmetry and  which is in good agreement with the PDG average value of $(0.72 \pm 0.10) \times 10^{-6}$~\cite{ParticleDataGroup:2024cfk}.





\section{\label{sectionV}Summary} 

In this work, we have calculated the factorizable contributions to the nonleptonic decays of $B$ mesons using light-cone sum rules. The results of the branching ratios for $B^0 \to \pi^+\pi^-$, $B^+ \to \pi^+\pi^0$, $B^0 \to K^+\pi^-$, and $B^0 \to K^+K^-$ are consistent with experimental measurements within the uncertainties, demonstrating the effectiveness of the LCSR method. Additionally, using the branching ratios for $B^0 \to K^+\pi^-$, we estimate the branching ratio for the decay channel $B_s^0 \to \pi^+\pi^-$, which is close to the average result given by the PDG.

Based on the experimentally measured branching ratios, we have derived the ratio of the CKM matrix element combinations $|V_{ub}V_{us}^\ast|/|V_{ub}V_{ud}^\ast| \approx 0.18$, which is smaller than the Wolfenstein parameter $\lambda \approx 0.23$. This discrepancy is primarily attributed to the different decay dynamics between the tree-level dominated $B^0 \to \pi^+\pi^-$ and the penguin-dominated $B^0 \to K^+\pi^-$ decays, while we include only the tree-level contributions for both decay channels. To improve the precision for $B^0 \to K^+\pi^-$ decay, the penguin diagram is needed to be included.

However, the branching ratios for $B \to \pi\pi$ were slightly smaller than the experimental values, which may be attributed to the neglect of higher order QCD corrections and non-factorizable contributions. To improve the results, it is essential to incorporate $\mathcal{O}(\alpha_s)$ loop corrections, take account of contributions from higher twist distribution amplitudes and three-particle light-cone distribution amplitudes of pseudoscalar mesons as additional non-perturbative inputs. With these improvements, the LCSR method is expected to play a more significant role in heavy flavor physics.

In the future, combining more experimental data and theoretical progress will allow for a deeper understanding of the decay mechanisms of $B$ meson, hereby providing a stronger support for the Standard Model or exploring possible new physics.

\appendix
\renewcommand{\thesection}{}
\section{Distribution amplitudes of pion and kaon}\label{DAs}

In this appendix, we present the Gegenbauer expansions of distribution amplitudes for $\pi$ and $K$ mesons. Eq.~(\ref{twist2}) gives the twist-2 distribution amplitudes of pion and kaon, while Eq.~(\ref{twist3pion}) and Eq.~(\ref{twist3K}) correspond to the twist-3, respectively.
 \begin{equation}\label{twist2}
	\varphi_{2;\pi(K)}(u,\mu)=6 u(1-u)[1+\sum_{n=1}^4 a_n^{\pi(K)}(\mu) C_n^{3 / 2}(2 u-1)],
 \end{equation}
 \begin{equation}\label{twist3pion}
	\varphi_{3;\pi}^{\mathcal{P}}(u,\mu)=1+B_2(\mu)\frac{1}{2}[3(u-\bar{u})^2-1]
	+B_4(\mu)\frac{1}{8}[35(u-\bar{u})^4-30(u-\bar{u})^2+3],
 \end{equation}
 \begin{equation}\label{twist3K}
	\begin{aligned}
		\varphi_{3;K}^{\mathcal{P}}(u,\mu)=&1+[30 \eta_3^K(\mu)-\frac{5}{2}] C_2^{1 / 2}(2 u-1)\\
		&+[-3\eta_3^K(\mu) \omega_3^K(\mu)-\frac{27}{20} \rho_3^{K}(\mu)^2-\frac{81}{10} \rho_3^K(\mu)^2a_2^K(\mu)] C_4^{1 / 2}(2 u-1),
	\end{aligned}
 \end{equation}
where $\bar{u}=1-u$.

\begin{acknowledgments}
This work was supported in part by National Natural Science Foundation of China under Grant No.11365018 and No.11375240.
\end{acknowledgments}

\bibliographystyle{apsrev4-1}
\bibliography{apssamp}

\begin{thebibliography}{42}%
\makeatletter
\providecommand \@ifxundefined [1]{%
 \@ifx{#1\undefined}
}%
\providecommand \@ifnum [1]{%
 \ifnum #1\expandafter \@firstoftwo
 \else \expandafter \@secondoftwo
 \fi
}%
\providecommand \@ifx [1]{%
 \ifx #1\expandafter \@firstoftwo
 \else \expandafter \@secondoftwo
 \fi
}%
\providecommand \natexlab [1]{#1}%
\providecommand \enquote  [1]{``#1''}%
\providecommand \bibnamefont  [1]{#1}%
\providecommand \bibfnamefont [1]{#1}%
\providecommand \citenamefont [1]{#1}%
\providecommand \href@noop [0]{\@secondoftwo}%
\providecommand \href [0]{\begingroup \@sanitize@url \@href}%
\providecommand \@href[1]{\@@startlink{#1}\@@href}%
\providecommand \@@href[1]{\endgroup#1\@@endlink}%
\providecommand \@sanitize@url [0]{\catcode `\\12\catcode `\$12\catcode
  `\&12\catcode `\#12\catcode `\^12\catcode `\_12\catcode `\%12\relax}%
\providecommand \@@startlink[1]{}%
\providecommand \@@endlink[0]{}%
\providecommand \url  [0]{\begingroup\@sanitize@url \@url }%
\providecommand \@url [1]{\endgroup\@href {#1}{\urlprefix }}%
\providecommand \urlprefix  [0]{URL }%
\providecommand \Eprint [0]{\href }%
\providecommand \doibase [0]{http://dx.doi.org/}%
\providecommand \selectlanguage [0]{\@gobble}%
\providecommand \bibinfo  [0]{\@secondoftwo}%
\providecommand \bibfield  [0]{\@secondoftwo}%
\providecommand \translation [1]{[#1]}%
\providecommand \BibitemOpen [0]{}%
\providecommand \bibitemStop [0]{}%
\providecommand \bibitemNoStop [0]{.\EOS\space}%
\providecommand \EOS [0]{\spacefactor3000\relax}%
\providecommand \BibitemShut  [1]{\csname bibitem#1\endcsname}%
\let\auto@bib@innerbib\@empty
\bibitem [{\citenamefont {Aguilar}\ \emph {et~al.}(2025)\citenamefont
  {Aguilar}, \citenamefont {Rend\'on},\ and\ \citenamefont
  {Roig}}]{Aguilar:2024ybr}%
  \BibitemOpen
  \bibfield  {author} {\bibinfo {author} {\bibfnamefont {D.~A.~L.}\
  \bibnamefont {Aguilar}}, \bibinfo {author} {\bibfnamefont {J.}~\bibnamefont
  {Rend\'on}}, \ and\ \bibinfo {author} {\bibfnamefont {P.}~\bibnamefont
  {Roig}},\ }\href {\doibase 10.1007/JHEP01(2025)105} {\bibfield  {journal}
  {\bibinfo  {journal} {JHEP}\ }\textbf {\bibinfo {volume} {01}},\ \bibinfo
  {pages} {105} (\bibinfo {year} {2025})},\ \Eprint
  {http://arxiv.org/abs/2409.05588} {arXiv:2409.05588 [hep-ph]} \BibitemShut
  {NoStop}%
\bibitem [{\citenamefont {Vale~Silva}\ \emph {et~al.}(2024)\citenamefont
  {Vale~Silva}, \citenamefont {Pich},\ and\ \citenamefont
  {Solomonidi}}]{ValeSilva:2024vmv}%
  \BibitemOpen
  \bibfield  {author} {\bibinfo {author} {\bibfnamefont {L.}~\bibnamefont
  {Vale~Silva}}, \bibinfo {author} {\bibfnamefont {A.}~\bibnamefont {Pich}}, \
  and\ \bibinfo {author} {\bibfnamefont {E.}~\bibnamefont {Solomonidi}},\
  }\href {\doibase 10.1142/S0217751X2442003X} {\bibfield  {journal} {\bibinfo
  {journal} {Int. J. Mod. Phys. A}\ }\textbf {\bibinfo {volume} {39}},\
  \bibinfo {pages} {2442003} (\bibinfo {year} {2024})}\BibitemShut {NoStop}%
\bibitem [{\citenamefont {Aaij}\ \emph {et~al.}(2015)\citenamefont {Aaij} \emph
  {et~al.}}]{LHCb:2014iah}%
  \BibitemOpen
  \bibfield  {author} {\bibinfo {author} {\bibfnamefont {R.}~\bibnamefont
  {Aaij}} \emph {et~al.} (\bibinfo {collaboration} {LHCb}),\ }\href {\doibase
  10.1103/PhysRevLett.114.041801} {\bibfield  {journal} {\bibinfo  {journal}
  {Phys. Rev. Lett.}\ }\textbf {\bibinfo {volume} {114}},\ \bibinfo {pages}
  {041801} (\bibinfo {year} {2015})},\ \Eprint {http://arxiv.org/abs/1411.3104}
  {arXiv:1411.3104 [hep-ex]} \BibitemShut {NoStop}%
\bibitem [{\citenamefont {Saito}\ \emph {et~al.}(2015)\citenamefont {Saito}
  \emph {et~al.}}]{Belle:2014nmp}%
  \BibitemOpen
  \bibfield  {author} {\bibinfo {author} {\bibfnamefont {T.}~\bibnamefont
  {Saito}} \emph {et~al.} (\bibinfo {collaboration} {Belle}),\ }\href {\doibase
  10.1103/PhysRevD.91.052004} {\bibfield  {journal} {\bibinfo  {journal} {Phys.
  Rev. D}\ }\textbf {\bibinfo {volume} {91}},\ \bibinfo {pages} {052004}
  (\bibinfo {year} {2015})},\ \Eprint {http://arxiv.org/abs/1411.7198}
  {arXiv:1411.7198 [hep-ex]} \BibitemShut {NoStop}%
\bibitem [{\citenamefont {Qi}\ \emph {et~al.}(2019)\citenamefont {Qi},
  \citenamefont {Wang}, \citenamefont {Guo}, \citenamefont {Zhang},\ and\
  \citenamefont {Wang}}]{Qi:2018syl}%
  \BibitemOpen
  \bibfield  {author} {\bibinfo {author} {\bibfnamefont {J.-J.}\ \bibnamefont
  {Qi}}, \bibinfo {author} {\bibfnamefont {Z.-Y.}\ \bibnamefont {Wang}},
  \bibinfo {author} {\bibfnamefont {X.-H.}\ \bibnamefont {Guo}}, \bibinfo
  {author} {\bibfnamefont {Z.-H.}\ \bibnamefont {Zhang}}, \ and\ \bibinfo
  {author} {\bibfnamefont {C.}~\bibnamefont {Wang}},\ }\href {\doibase
  10.1103/PhysRevD.99.076010} {\bibfield  {journal} {\bibinfo  {journal} {Phys.
  Rev. D}\ }\textbf {\bibinfo {volume} {99}},\ \bibinfo {pages} {076010}
  (\bibinfo {year} {2019})},\ \Eprint {http://arxiv.org/abs/1811.02167}
  {arXiv:1811.02167 [hep-ph]} \BibitemShut {NoStop}%
\bibitem [{\citenamefont {Brambilla}\ \emph {et~al.}(2011)\citenamefont
  {Brambilla} \emph {et~al.}}]{Brambilla:2010cs}%
  \BibitemOpen
  \bibfield  {author} {\bibinfo {author} {\bibfnamefont {N.}~\bibnamefont
  {Brambilla}} \emph {et~al.},\ }\href {\doibase
  10.1140/epjc/s10052-010-1534-9} {\bibfield  {journal} {\bibinfo  {journal}
  {Eur. Phys. J. C}\ }\textbf {\bibinfo {volume} {71}},\ \bibinfo {pages}
  {1534} (\bibinfo {year} {2011})},\ \Eprint {http://arxiv.org/abs/1010.5827}
  {arXiv:1010.5827 [hep-ph]} \BibitemShut {NoStop}%
\bibitem [{\citenamefont {Eichten}\ and\ \citenamefont
  {Quigg}(2017)}]{Eichten:2017ffp}%
  \BibitemOpen
  \bibfield  {author} {\bibinfo {author} {\bibfnamefont {E.~J.}\ \bibnamefont
  {Eichten}}\ and\ \bibinfo {author} {\bibfnamefont {C.}~\bibnamefont
  {Quigg}},\ }\href {\doibase 10.1103/PhysRevLett.119.202002} {\bibfield
  {journal} {\bibinfo  {journal} {Phys. Rev. Lett.}\ }\textbf {\bibinfo
  {volume} {119}},\ \bibinfo {pages} {202002} (\bibinfo {year} {2017})},\
  \Eprint {http://arxiv.org/abs/1707.09575} {arXiv:1707.09575 [hep-ph]}
  \BibitemShut {NoStop}%
\bibitem [{\citenamefont {Aaij}\ \emph {et~al.}(2018)\citenamefont {Aaij} \emph
  {et~al.}}]{LHCb:2018vuc}%
  \BibitemOpen
  \bibfield  {author} {\bibinfo {author} {\bibfnamefont {R.}~\bibnamefont
  {Aaij}} \emph {et~al.} (\bibinfo {collaboration} {LHCb}),\ }\href {\doibase
  10.1103/PhysRevLett.121.072002} {\bibfield  {journal} {\bibinfo  {journal}
  {Phys. Rev. Lett.}\ }\textbf {\bibinfo {volume} {121}},\ \bibinfo {pages}
  {072002} (\bibinfo {year} {2018})},\ \Eprint
  {http://arxiv.org/abs/1805.09418} {arXiv:1805.09418 [hep-ex]} \BibitemShut
  {NoStop}%
\bibitem [{\citenamefont {Abe}\ \emph {et~al.}(2001)\citenamefont {Abe} \emph
  {et~al.}}]{Belle:2001zzw}%
  \BibitemOpen
  \bibfield  {author} {\bibinfo {author} {\bibfnamefont {K.}~\bibnamefont
  {Abe}} \emph {et~al.} (\bibinfo {collaboration} {Belle}),\ }\href {\doibase
  10.1103/PhysRevLett.87.091802} {\bibfield  {journal} {\bibinfo  {journal}
  {Phys. Rev. Lett.}\ }\textbf {\bibinfo {volume} {87}},\ \bibinfo {pages}
  {091802} (\bibinfo {year} {2001})},\ \Eprint
  {http://arxiv.org/abs/hep-ex/0107061} {arXiv:hep-ex/0107061} \BibitemShut
  {NoStop}%
\bibitem [{\citenamefont {Aubert}\ \emph {et~al.}(2002)\citenamefont {Aubert}
  \emph {et~al.}}]{BaBar:2002kla}%
  \BibitemOpen
  \bibfield  {author} {\bibinfo {author} {\bibfnamefont {B.}~\bibnamefont
  {Aubert}} \emph {et~al.} (\bibinfo {collaboration} {BaBar}),\ }\href
  {\doibase 10.1103/PhysRevLett.89.201802} {\bibfield  {journal} {\bibinfo
  {journal} {Phys. Rev. Lett.}\ }\textbf {\bibinfo {volume} {89}},\ \bibinfo
  {pages} {201802} (\bibinfo {year} {2002})},\ \Eprint
  {http://arxiv.org/abs/hep-ex/0207042} {arXiv:hep-ex/0207042} \BibitemShut
  {NoStop}%
\bibitem [{\citenamefont {Aaij}\ \emph {et~al.}(2019)\citenamefont {Aaij} \emph
  {et~al.}}]{LHCb:2019hro}%
  \BibitemOpen
  \bibfield  {author} {\bibinfo {author} {\bibfnamefont {R.}~\bibnamefont
  {Aaij}} \emph {et~al.} (\bibinfo {collaboration} {LHCb}),\ }\href {\doibase
  10.1103/PhysRevLett.122.211803} {\bibfield  {journal} {\bibinfo  {journal}
  {Phys. Rev. Lett.}\ }\textbf {\bibinfo {volume} {122}},\ \bibinfo {pages}
  {211803} (\bibinfo {year} {2019})},\ \Eprint
  {http://arxiv.org/abs/1903.08726} {arXiv:1903.08726 [hep-ex]} \BibitemShut
  {NoStop}%
\bibitem [{\citenamefont {Wirbel}\ \emph {et~al.}(1985)\citenamefont {Wirbel},
  \citenamefont {Stech},\ and\ \citenamefont {Bauer}}]{Wirbel:1985ji}%
  \BibitemOpen
  \bibfield  {author} {\bibinfo {author} {\bibfnamefont {M.}~\bibnamefont
  {Wirbel}}, \bibinfo {author} {\bibfnamefont {B.}~\bibnamefont {Stech}}, \
  and\ \bibinfo {author} {\bibfnamefont {M.}~\bibnamefont {Bauer}},\ }\href
  {\doibase 10.1007/BF01560299} {\bibfield  {journal} {\bibinfo  {journal} {Z.
  Phys. C}\ }\textbf {\bibinfo {volume} {29}},\ \bibinfo {pages} {637}
  (\bibinfo {year} {1985})}\BibitemShut {NoStop}%
\bibitem [{\citenamefont {Bauer}\ and\ \citenamefont
  {Stech}(1985)}]{Bauer:1984zv}%
  \BibitemOpen
  \bibfield  {author} {\bibinfo {author} {\bibfnamefont {M.}~\bibnamefont
  {Bauer}}\ and\ \bibinfo {author} {\bibfnamefont {B.}~\bibnamefont {Stech}},\
  }\href {\doibase 10.1016/0370-2693(85)90515-5} {\bibfield  {journal}
  {\bibinfo  {journal} {Phys. Lett. B}\ }\textbf {\bibinfo {volume} {152}},\
  \bibinfo {pages} {380} (\bibinfo {year} {1985})}\BibitemShut {NoStop}%
\bibitem [{\citenamefont {Kramer}\ and\ \citenamefont
  {Palmer}(1992)}]{Kramer:1992xr}%
  \BibitemOpen
  \bibfield  {author} {\bibinfo {author} {\bibfnamefont {G.}~\bibnamefont
  {Kramer}}\ and\ \bibinfo {author} {\bibfnamefont {W.~F.}\ \bibnamefont
  {Palmer}},\ }\href {\doibase 10.1103/PhysRevD.46.3197} {\bibfield  {journal}
  {\bibinfo  {journal} {Phys. Rev. D}\ }\textbf {\bibinfo {volume} {46}},\
  \bibinfo {pages} {3197} (\bibinfo {year} {1992})}\BibitemShut {NoStop}%
\bibitem [{\citenamefont {Du}\ and\ \citenamefont {Yang}(1995)}]{Du:1995qy}%
  \BibitemOpen
  \bibfield  {author} {\bibinfo {author} {\bibfnamefont {D.-s.}\ \bibnamefont
  {Du}}\ and\ \bibinfo {author} {\bibfnamefont {M.-z.}\ \bibnamefont {Yang}},\
  }\href {\doibase 10.1016/0370-2693(95)00959-O} {\bibfield  {journal}
  {\bibinfo  {journal} {Phys. Lett. B}\ }\textbf {\bibinfo {volume} {358}},\
  \bibinfo {pages} {123} (\bibinfo {year} {1995})},\ \Eprint
  {http://arxiv.org/abs/hep-ph/9503278} {arXiv:hep-ph/9503278} \BibitemShut
  {NoStop}%
\bibitem [{\citenamefont {Beneke}\ \emph {et~al.}(2000)\citenamefont {Beneke},
  \citenamefont {Buchalla}, \citenamefont {Neubert},\ and\ \citenamefont
  {Sachrajda}}]{Beneke:2000ry}%
  \BibitemOpen
  \bibfield  {author} {\bibinfo {author} {\bibfnamefont {M.}~\bibnamefont
  {Beneke}}, \bibinfo {author} {\bibfnamefont {G.}~\bibnamefont {Buchalla}},
  \bibinfo {author} {\bibfnamefont {M.}~\bibnamefont {Neubert}}, \ and\
  \bibinfo {author} {\bibfnamefont {C.~T.}\ \bibnamefont {Sachrajda}},\ }\href
  {\doibase 10.1016/S0550-3213(00)00559-9} {\bibfield  {journal} {\bibinfo
  {journal} {Nucl. Phys. B}\ }\textbf {\bibinfo {volume} {591}},\ \bibinfo
  {pages} {313} (\bibinfo {year} {2000})},\ \Eprint
  {http://arxiv.org/abs/hep-ph/0006124} {arXiv:hep-ph/0006124} \BibitemShut
  {NoStop}%
\bibitem [{\citenamefont {Kivel}(2022)}]{Kivel:2021uzl}%
  \BibitemOpen
  \bibfield  {author} {\bibinfo {author} {\bibfnamefont {N.}~\bibnamefont
  {Kivel}},\ }\href {\doibase 10.1140/epja/s10050-021-00626-1} {\bibfield
  {journal} {\bibinfo  {journal} {Eur. Phys. J. A}\ }\textbf {\bibinfo {volume}
  {58}},\ \bibinfo {pages} {26} (\bibinfo {year} {2022})},\ \Eprint
  {http://arxiv.org/abs/2109.05847} {arXiv:2109.05847 [hep-ph]} \BibitemShut
  {NoStop}%
\bibitem [{\citenamefont {Li}\ and\ \citenamefont {Yu}(1995)}]{Li:1994cka}%
  \BibitemOpen
  \bibfield  {author} {\bibinfo {author} {\bibfnamefont {H.-n.}\ \bibnamefont
  {Li}}\ and\ \bibinfo {author} {\bibfnamefont {H.-L.}\ \bibnamefont {Yu}},\
  }\href {\doibase 10.1103/PhysRevLett.74.4388} {\bibfield  {journal} {\bibinfo
   {journal} {Phys. Rev. Lett.}\ }\textbf {\bibinfo {volume} {74}},\ \bibinfo
  {pages} {4388} (\bibinfo {year} {1995})},\ \Eprint
  {http://arxiv.org/abs/hep-ph/9409313} {arXiv:hep-ph/9409313} \BibitemShut
  {NoStop}%
\bibitem [{\citenamefont {Li}(2003)}]{Li:2003yj}%
  \BibitemOpen
  \bibfield  {author} {\bibinfo {author} {\bibfnamefont {H.-n.}\ \bibnamefont
  {Li}},\ }\href {\doibase 10.1016/S0146-6410(03)90013-5} {\bibfield  {journal}
  {\bibinfo  {journal} {Prog. Part. Nucl. Phys.}\ }\textbf {\bibinfo {volume}
  {51}},\ \bibinfo {pages} {85} (\bibinfo {year} {2003})},\ \Eprint
  {http://arxiv.org/abs/hep-ph/0303116} {arXiv:hep-ph/0303116} \BibitemShut
  {NoStop}%
\bibitem [{\citenamefont {Li}\ \emph {et~al.}(2022)\citenamefont {Li},
  \citenamefont {Zhao}, \citenamefont {Sun},\ and\ \citenamefont
  {Zou}}]{Li:2022mtc}%
  \BibitemOpen
  \bibfield  {author} {\bibinfo {author} {\bibfnamefont {Y.}~\bibnamefont
  {Li}}, \bibinfo {author} {\bibfnamefont {G.-H.}\ \bibnamefont {Zhao}},
  \bibinfo {author} {\bibfnamefont {Y.-J.}\ \bibnamefont {Sun}}, \ and\
  \bibinfo {author} {\bibfnamefont {Z.-T.}\ \bibnamefont {Zou}},\ }\href
  {\doibase 10.1103/PhysRevD.106.093009} {\bibfield  {journal} {\bibinfo
  {journal} {Phys. Rev. D}\ }\textbf {\bibinfo {volume} {106}},\ \bibinfo
  {pages} {093009} (\bibinfo {year} {2022})},\ \Eprint
  {http://arxiv.org/abs/2209.13389} {arXiv:2209.13389 [hep-ph]} \BibitemShut
  {NoStop}%
\bibitem [{\citenamefont {Bauer}\ \emph {et~al.}(2001)\citenamefont {Bauer},
  \citenamefont {Pirjol},\ and\ \citenamefont {Stewart}}]{Bauer:2001cu}%
  \BibitemOpen
  \bibfield  {author} {\bibinfo {author} {\bibfnamefont {C.~W.}\ \bibnamefont
  {Bauer}}, \bibinfo {author} {\bibfnamefont {D.}~\bibnamefont {Pirjol}}, \
  and\ \bibinfo {author} {\bibfnamefont {I.~W.}\ \bibnamefont {Stewart}},\
  }\href {\doibase 10.1103/PhysRevLett.87.201806} {\bibfield  {journal}
  {\bibinfo  {journal} {Phys. Rev. Lett.}\ }\textbf {\bibinfo {volume} {87}},\
  \bibinfo {pages} {201806} (\bibinfo {year} {2001})},\ \Eprint
  {http://arxiv.org/abs/hep-ph/0107002} {arXiv:hep-ph/0107002} \BibitemShut
  {NoStop}%
\bibitem [{\citenamefont {Bauer}\ and\ \citenamefont
  {Stewart}(2001)}]{Bauer:2001ct}%
  \BibitemOpen
  \bibfield  {author} {\bibinfo {author} {\bibfnamefont {C.~W.}\ \bibnamefont
  {Bauer}}\ and\ \bibinfo {author} {\bibfnamefont {I.~W.}\ \bibnamefont
  {Stewart}},\ }\href {\doibase 10.1016/S0370-2693(01)00902-9} {\bibfield
  {journal} {\bibinfo  {journal} {Phys. Lett. B}\ }\textbf {\bibinfo {volume}
  {516}},\ \bibinfo {pages} {134} (\bibinfo {year} {2001})},\ \Eprint
  {http://arxiv.org/abs/hep-ph/0107001} {arXiv:hep-ph/0107001} \BibitemShut
  {NoStop}%
\bibitem [{\citenamefont {Bauer}\ \emph {et~al.}(2003)\citenamefont {Bauer},
  \citenamefont {Pirjol},\ and\ \citenamefont {Stewart}}]{Bauer:2002aj}%
  \BibitemOpen
  \bibfield  {author} {\bibinfo {author} {\bibfnamefont {C.~W.}\ \bibnamefont
  {Bauer}}, \bibinfo {author} {\bibfnamefont {D.}~\bibnamefont {Pirjol}}, \
  and\ \bibinfo {author} {\bibfnamefont {I.~W.}\ \bibnamefont {Stewart}},\
  }\href {\doibase 10.1103/PhysRevD.67.071502} {\bibfield  {journal} {\bibinfo
  {journal} {Phys. Rev. D}\ }\textbf {\bibinfo {volume} {67}},\ \bibinfo
  {pages} {071502} (\bibinfo {year} {2003})},\ \Eprint
  {http://arxiv.org/abs/hep-ph/0211069} {arXiv:hep-ph/0211069} \BibitemShut
  {NoStop}%
\bibitem [{\citenamefont {Khodjamirian}(2001)}]{Khodjamirian:2000mi}%
  \BibitemOpen
  \bibfield  {author} {\bibinfo {author} {\bibfnamefont {A.}~\bibnamefont
  {Khodjamirian}},\ }\href {\doibase 10.1016/S0550-3213(01)00194-8} {\bibfield
  {journal} {\bibinfo  {journal} {Nucl. Phys. B}\ }\textbf {\bibinfo {volume}
  {605}},\ \bibinfo {pages} {558} (\bibinfo {year} {2001})},\ \Eprint
  {http://arxiv.org/abs/hep-ph/0012271} {arXiv:hep-ph/0012271} \BibitemShut
  {NoStop}%
\bibitem [{\citenamefont {Wu}\ \emph {et~al.}(2002)\citenamefont {Wu},
  \citenamefont {Li}, \citenamefont {Cui},\ and\ \citenamefont
  {Huang}}]{Wu:2002csa}%
  \BibitemOpen
  \bibfield  {author} {\bibinfo {author} {\bibfnamefont {X.-Y.}\ \bibnamefont
  {Wu}}, \bibinfo {author} {\bibfnamefont {Z.-H.}\ \bibnamefont {Li}}, \bibinfo
  {author} {\bibfnamefont {J.-Y.}\ \bibnamefont {Cui}}, \ and\ \bibinfo
  {author} {\bibfnamefont {T.}~\bibnamefont {Huang}},\ }\href {\doibase
  10.1088/0256-307X/19/11/309} {\bibfield  {journal} {\bibinfo  {journal}
  {Chin. Phys. Lett.}\ }\textbf {\bibinfo {volume} {19}},\ \bibinfo {pages}
  {1596} (\bibinfo {year} {2002})}\BibitemShut {NoStop}%
\bibitem [{\citenamefont {Piscopo}\ and\ \citenamefont
  {Rusov}(2023)}]{Piscopo:2023opf}%
  \BibitemOpen
  \bibfield  {author} {\bibinfo {author} {\bibfnamefont {M.~L.}\ \bibnamefont
  {Piscopo}}\ and\ \bibinfo {author} {\bibfnamefont {A.~V.}\ \bibnamefont
  {Rusov}},\ }\href {\doibase 10.1007/JHEP10(2023)180} {\bibfield  {journal}
  {\bibinfo  {journal} {JHEP}\ }\textbf {\bibinfo {volume} {10}},\ \bibinfo
  {pages} {180} (\bibinfo {year} {2023})},\ \Eprint
  {http://arxiv.org/abs/2307.07594} {arXiv:2307.07594 [hep-ph]} \BibitemShut
  {NoStop}%
\bibitem [{\citenamefont {Khodjamirian}\ \emph
  {et~al.}(2003{\natexlab{a}})\citenamefont {Khodjamirian}, \citenamefont
  {Mannel},\ and\ \citenamefont {Urban}}]{Khodjamirian:2002pk}%
  \BibitemOpen
  \bibfield  {author} {\bibinfo {author} {\bibfnamefont {A.}~\bibnamefont
  {Khodjamirian}}, \bibinfo {author} {\bibfnamefont {T.}~\bibnamefont
  {Mannel}}, \ and\ \bibinfo {author} {\bibfnamefont {P.}~\bibnamefont
  {Urban}},\ }\href {\doibase 10.1103/PhysRevD.67.054027} {\bibfield  {journal}
  {\bibinfo  {journal} {Phys. Rev. D}\ }\textbf {\bibinfo {volume} {67}},\
  \bibinfo {pages} {054027} (\bibinfo {year} {2003}{\natexlab{a}})},\ \Eprint
  {http://arxiv.org/abs/hep-ph/0210378} {arXiv:hep-ph/0210378} \BibitemShut
  {NoStop}%
\bibitem [{\citenamefont {Khodjamirian}\ \emph
  {et~al.}(2003{\natexlab{b}})\citenamefont {Khodjamirian}, \citenamefont
  {Mannel},\ and\ \citenamefont {Melic}}]{Khodjamirian:2003eq}%
  \BibitemOpen
  \bibfield  {author} {\bibinfo {author} {\bibfnamefont {A.}~\bibnamefont
  {Khodjamirian}}, \bibinfo {author} {\bibfnamefont {T.}~\bibnamefont
  {Mannel}}, \ and\ \bibinfo {author} {\bibfnamefont {B.}~\bibnamefont
  {Melic}},\ }\href {\doibase 10.1016/j.physletb.2003.08.012} {\bibfield
  {journal} {\bibinfo  {journal} {Phys. Lett. B}\ }\textbf {\bibinfo {volume}
  {571}},\ \bibinfo {pages} {75} (\bibinfo {year} {2003}{\natexlab{b}})},\
  \Eprint {http://arxiv.org/abs/hep-ph/0304179} {arXiv:hep-ph/0304179}
  \BibitemShut {NoStop}%
\bibitem [{\citenamefont {Khodjamirian}\ \emph {et~al.}(2005)\citenamefont
  {Khodjamirian}, \citenamefont {Mannel}, \citenamefont {Melcher},\ and\
  \citenamefont {Melic}}]{Khodjamirian:2005wn}%
  \BibitemOpen
  \bibfield  {author} {\bibinfo {author} {\bibfnamefont {A.}~\bibnamefont
  {Khodjamirian}}, \bibinfo {author} {\bibfnamefont {T.}~\bibnamefont
  {Mannel}}, \bibinfo {author} {\bibfnamefont {M.}~\bibnamefont {Melcher}}, \
  and\ \bibinfo {author} {\bibfnamefont {B.}~\bibnamefont {Melic}},\ }\href
  {\doibase 10.1103/PhysRevD.72.094012} {\bibfield  {journal} {\bibinfo
  {journal} {Phys. Rev. D}\ }\textbf {\bibinfo {volume} {72}},\ \bibinfo
  {pages} {094012} (\bibinfo {year} {2005})},\ \Eprint
  {http://arxiv.org/abs/hep-ph/0509049} {arXiv:hep-ph/0509049} \BibitemShut
  {NoStop}%
\bibitem [{\citenamefont {Shi}\ and\ \citenamefont {Zhao}(2024)}]{Shi:2024plf}%
  \BibitemOpen
  \bibfield  {author} {\bibinfo {author} {\bibfnamefont {Y.-J.}\ \bibnamefont
  {Shi}}\ and\ \bibinfo {author} {\bibfnamefont {Z.-X.}\ \bibnamefont {Zhao}},\
  }\href {\doibase 10.1103/PhysRevD.110.096015} {\bibfield  {journal} {\bibinfo
   {journal} {Phys. Rev. D}\ }\textbf {\bibinfo {volume} {110}},\ \bibinfo
  {pages} {096015} (\bibinfo {year} {2024})},\ \Eprint
  {http://arxiv.org/abs/2407.07431} {arXiv:2407.07431 [hep-ph]} \BibitemShut
  {NoStop}%
\bibitem [{\citenamefont {Buchalla}\ \emph {et~al.}(1996)\citenamefont
  {Buchalla}, \citenamefont {Buras},\ and\ \citenamefont
  {Lautenbacher}}]{Buchalla:1995vs}%
  \BibitemOpen
  \bibfield  {author} {\bibinfo {author} {\bibfnamefont {G.}~\bibnamefont
  {Buchalla}}, \bibinfo {author} {\bibfnamefont {A.~J.}\ \bibnamefont {Buras}},
  \ and\ \bibinfo {author} {\bibfnamefont {M.~E.}\ \bibnamefont
  {Lautenbacher}},\ }\href {\doibase 10.1103/RevModPhys.68.1125} {\bibfield
  {journal} {\bibinfo  {journal} {Rev. Mod. Phys.}\ }\textbf {\bibinfo {volume}
  {68}},\ \bibinfo {pages} {1125} (\bibinfo {year} {1996})},\ \Eprint
  {http://arxiv.org/abs/hep-ph/9512380} {arXiv:hep-ph/9512380} \BibitemShut
  {NoStop}%
\bibitem [{\citenamefont {Buras}\ and\ \citenamefont
  {Silvestrini}(2000)}]{Buras:1998ra}%
  \BibitemOpen
  \bibfield  {author} {\bibinfo {author} {\bibfnamefont {A.~J.}\ \bibnamefont
  {Buras}}\ and\ \bibinfo {author} {\bibfnamefont {L.}~\bibnamefont
  {Silvestrini}},\ }\href {\doibase 10.1016/S0550-3213(99)00712-9} {\bibfield
  {journal} {\bibinfo  {journal} {Nucl. Phys. B}\ }\textbf {\bibinfo {volume}
  {569}},\ \bibinfo {pages} {3} (\bibinfo {year} {2000})},\ \Eprint
  {http://arxiv.org/abs/hep-ph/9812392} {arXiv:hep-ph/9812392} \BibitemShut
  {NoStop}%
\bibitem [{\citenamefont {Belyaev}\ \emph {et~al.}(1995)\citenamefont
  {Belyaev}, \citenamefont {Braun}, \citenamefont {Khodjamirian},\ and\
  \citenamefont {Ruckl}}]{Belyaev:1994zk}%
  \BibitemOpen
  \bibfield  {author} {\bibinfo {author} {\bibfnamefont {V.~M.}\ \bibnamefont
  {Belyaev}}, \bibinfo {author} {\bibfnamefont {V.~M.}\ \bibnamefont {Braun}},
  \bibinfo {author} {\bibfnamefont {A.}~\bibnamefont {Khodjamirian}}, \ and\
  \bibinfo {author} {\bibfnamefont {R.}~\bibnamefont {Ruckl}},\ }\href
  {\doibase 10.1103/PhysRevD.51.6177} {\bibfield  {journal} {\bibinfo
  {journal} {Phys. Rev. D}\ }\textbf {\bibinfo {volume} {51}},\ \bibinfo
  {pages} {6177} (\bibinfo {year} {1995})},\ \Eprint
  {http://arxiv.org/abs/hep-ph/9410280} {arXiv:hep-ph/9410280} \BibitemShut
  {NoStop}%
\bibitem [{\citenamefont {Navas}\ \emph {et~al.}(2024)\citenamefont {Navas}
  \emph {et~al.}}]{ParticleDataGroup:2024cfk}%
  \BibitemOpen
  \bibfield  {author} {\bibinfo {author} {\bibfnamefont {S.}~\bibnamefont
  {Navas}} \emph {et~al.} (\bibinfo {collaboration} {Particle Data Group}),\
  }\href {\doibase 10.1103/PhysRevD.110.030001} {\bibfield  {journal} {\bibinfo
   {journal} {Phys. Rev. D}\ }\textbf {\bibinfo {volume} {110}},\ \bibinfo
  {pages} {030001} (\bibinfo {year} {2024})}\BibitemShut {NoStop}%
\bibitem [{\citenamefont {Arifi}\ \emph {et~al.}(2022)\citenamefont {Arifi},
  \citenamefont {Choi}, \citenamefont {ji},\ and\ \citenamefont
  {Oh}}]{Arifi:2022pal}%
  \BibitemOpen
  \bibfield  {author} {\bibinfo {author} {\bibfnamefont {A.~J.}\ \bibnamefont
  {Arifi}}, \bibinfo {author} {\bibfnamefont {H.-M.}\ \bibnamefont {Choi}},
  \bibinfo {author} {\bibfnamefont {C.-R.}\ \bibnamefont {ji}}, \ and\ \bibinfo
  {author} {\bibfnamefont {Y.}~\bibnamefont {Oh}},\ }\href {\doibase
  10.1103/PhysRevD.106.014009} {\bibfield  {journal} {\bibinfo  {journal}
  {Phys. Rev. D}\ }\textbf {\bibinfo {volume} {106}},\ \bibinfo {pages}
  {014009} (\bibinfo {year} {2022})},\ \Eprint
  {http://arxiv.org/abs/2205.04075} {arXiv:2205.04075 [hep-ph]} \BibitemShut
  {NoStop}%
\bibitem [{\citenamefont {Ball}\ and\ \citenamefont
  {Zwicky}(2005)}]{Ball:2004ye}%
  \BibitemOpen
  \bibfield  {author} {\bibinfo {author} {\bibfnamefont {P.}~\bibnamefont
  {Ball}}\ and\ \bibinfo {author} {\bibfnamefont {R.}~\bibnamefont {Zwicky}},\
  }\href {\doibase 10.1103/PhysRevD.71.014015} {\bibfield  {journal} {\bibinfo
  {journal} {Phys. Rev. D}\ }\textbf {\bibinfo {volume} {71}},\ \bibinfo
  {pages} {014015} (\bibinfo {year} {2005})},\ \Eprint
  {http://arxiv.org/abs/hep-ph/0406232} {arXiv:hep-ph/0406232} \BibitemShut
  {NoStop}%
\bibitem [{\citenamefont {Chai}\ \emph {et~al.}(2022)\citenamefont {Chai},
  \citenamefont {Cheng}, \citenamefont {Ju}, \citenamefont {Yan}, \citenamefont
  {L\"u},\ and\ \citenamefont {Xiao}}]{Chai:2022ptk}%
  \BibitemOpen
  \bibfield  {author} {\bibinfo {author} {\bibfnamefont {J.}~\bibnamefont
  {Chai}}, \bibinfo {author} {\bibfnamefont {S.}~\bibnamefont {Cheng}},
  \bibinfo {author} {\bibfnamefont {Y.-h.}\ \bibnamefont {Ju}}, \bibinfo
  {author} {\bibfnamefont {D.-C.}\ \bibnamefont {Yan}}, \bibinfo {author}
  {\bibfnamefont {C.-D.}\ \bibnamefont {L\"u}}, \ and\ \bibinfo {author}
  {\bibfnamefont {Z.-J.}\ \bibnamefont {Xiao}},\ }\href {\doibase
  10.1088/1674-1137/ac88bd} {\bibfield  {journal} {\bibinfo  {journal} {Chin.
  Phys. C}\ }\textbf {\bibinfo {volume} {46}},\ \bibinfo {pages} {123103}
  (\bibinfo {year} {2022})},\ \Eprint {http://arxiv.org/abs/2207.04190}
  {arXiv:2207.04190 [hep-ph]} \BibitemShut {NoStop}%
\bibitem [{\citenamefont {Cheng}\ and\ \citenamefont
  {Chua}(2009)}]{Cheng:2009cn}%
  \BibitemOpen
  \bibfield  {author} {\bibinfo {author} {\bibfnamefont {H.-Y.}\ \bibnamefont
  {Cheng}}\ and\ \bibinfo {author} {\bibfnamefont {C.-K.}\ \bibnamefont
  {Chua}},\ }\href {\doibase 10.1103/PhysRevD.80.114008} {\bibfield  {journal}
  {\bibinfo  {journal} {Phys. Rev. D}\ }\textbf {\bibinfo {volume} {80}},\
  \bibinfo {pages} {114008} (\bibinfo {year} {2009})},\ \Eprint
  {http://arxiv.org/abs/0909.5229} {arXiv:0909.5229 [hep-ph]} \BibitemShut
  {NoStop}%
\bibitem [{\citenamefont {Adachi}\ \emph {et~al.}(2024)\citenamefont {Adachi}
  \emph {et~al.}}]{Belle-II:2023ksq}%
  \BibitemOpen
  \bibfield  {author} {\bibinfo {author} {\bibfnamefont {I.}~\bibnamefont
  {Adachi}} \emph {et~al.} (\bibinfo {collaboration} {Belle-II}),\ }\href
  {\doibase 10.1103/PhysRevD.109.012001} {\bibfield  {journal} {\bibinfo
  {journal} {Phys. Rev. D}\ }\textbf {\bibinfo {volume} {109}},\ \bibinfo
  {pages} {012001} (\bibinfo {year} {2024})},\ \Eprint
  {http://arxiv.org/abs/2310.06381} {arXiv:2310.06381 [hep-ex]} \BibitemShut
  {NoStop}%
\bibitem [{\citenamefont {Aaij}\ \emph {et~al.}(2017)\citenamefont {Aaij} \emph
  {et~al.}}]{LHCb:2016inp}%
  \BibitemOpen
  \bibfield  {author} {\bibinfo {author} {\bibfnamefont {R.}~\bibnamefont
  {Aaij}} \emph {et~al.} (\bibinfo {collaboration} {LHCb}),\ }\href {\doibase
  10.1103/PhysRevLett.118.081801} {\bibfield  {journal} {\bibinfo  {journal}
  {Phys. Rev. Lett.}\ }\textbf {\bibinfo {volume} {118}},\ \bibinfo {pages}
  {081801} (\bibinfo {year} {2017})},\ \Eprint
  {http://arxiv.org/abs/1610.08288} {arXiv:1610.08288 [hep-ex]} \BibitemShut
  {NoStop}%
\bibitem [{\citenamefont {Gronau}\ \emph {et~al.}(1994)\citenamefont {Gronau},
  \citenamefont {Hernandez}, \citenamefont {London},\ and\ \citenamefont
  {Rosner}}]{Gronau:1994rj}%
  \BibitemOpen
  \bibfield  {author} {\bibinfo {author} {\bibfnamefont {M.}~\bibnamefont
  {Gronau}}, \bibinfo {author} {\bibfnamefont {O.~F.}\ \bibnamefont
  {Hernandez}}, \bibinfo {author} {\bibfnamefont {D.}~\bibnamefont {London}}, \
  and\ \bibinfo {author} {\bibfnamefont {J.~L.}\ \bibnamefont {Rosner}},\
  }\href {\doibase 10.1103/PhysRevD.50.4529} {\bibfield  {journal} {\bibinfo
  {journal} {Phys. Rev. D}\ }\textbf {\bibinfo {volume} {50}},\ \bibinfo
  {pages} {4529} (\bibinfo {year} {1994})},\ \Eprint
  {http://arxiv.org/abs/hep-ph/9404283} {arXiv:hep-ph/9404283} \BibitemShut
  {NoStop}%
\bibitem [{\citenamefont {Aaij}\ \emph {et~al.}(2013)\citenamefont {Aaij} \emph
  {et~al.}}]{LHCb:2013vfg}%
  \BibitemOpen
  \bibfield  {author} {\bibinfo {author} {\bibfnamefont {R.}~\bibnamefont
  {Aaij}} \emph {et~al.} (\bibinfo {collaboration} {LHCb}),\ }\href {\doibase
  10.1007/JHEP04(2013)001} {\bibfield  {journal} {\bibinfo  {journal} {JHEP}\
  }\textbf {\bibinfo {volume} {04}},\ \bibinfo {pages} {001} (\bibinfo {year}
  {2013})},\ \Eprint {http://arxiv.org/abs/1301.5286} {arXiv:1301.5286
  [hep-ex]} \BibitemShut {NoStop}%
\end{thebibliography}%

\end{document}